\documentclass[aps,prl,reprint]{revtex4-1}
\usepackage{blindtext}
\usepackage{color}

\usepackage{graphicx}
\usepackage{subfigure}
\usepackage{dcolumn}
\usepackage{bm}
\usepackage{soul} 
\input epsf

\def\erf{{\rm erf}}
\def\i{{\rm i}}
\def\d{{\rm d}}
\def\exp{{\rm {exp}}}
\def\ln{{\rm {ln}}}

\def\Re{{{\rm Re}}}

\def\Sc{{{\rm Sc}}}
\def\ln{{\rm ln}}
\def\u{{\rm \bf u}}

\def\sinh{{\rm sinh}}

\newcommand\beq{\begin{equation}}
\newcommand\eeq{\end{equation}}

\def\i{{\rm \bf i}}

\begin{document}
\title{Linear stability analysis and direct numerical simulation of two layer channel flow}
\author{Kirti Chandra Sahu}
\email{ksahu@iith.ac.in}
\author{Rama Govindarajan$^{\dagger}$}
\affiliation{
Department of Chemical Engineering, Indian Institute of Technology Hyderabad, Sangareddy 502 285, Telangana, India\\
$^{\dagger}$TIFR Centre for Interdisciplinary Sciences, Tata Institute of Fundamental Research Narsingi, Hyderabad 500 075, India}

\begin{abstract}
We study the stability of two-fluid flow through a plane channel at Reynolds numbers of a hundred to a thousand in the linear and nonlinear regimes. The two fluids have the same density but different viscosities. The fluids, when miscible, are separated from each other by a mixed layer of small but finite thickness, across which viscosity changes from that of one fluid to that of the other. When immiscible, the interface is sharp. Our study spans a range of Schmidt numbers, viscosity ratios and location and thickness of the mixed layer.

A region of instability distinct from that of the Tollmien-Schlichting mode is obtained at moderate Reynolds numbers. We show that the overlap of the layer of viscosity-stratification with the critical layer of the dominant disturbance provides a mechanism for this instability. At very low values of diffusivity, the miscible flow behaves exactly like the immiscible in terms of stability characteristics. High levels of miscibility make the flow more stable. At intermediate levels of diffusivity however, in both linear and non-linear regimes, miscible flow can be more unstable than the corresponding immiscible flow without surface tension. This difference is greater when the thickness of the mixed layer is decreased, since the thinner the layer of viscosity stratification, the more unstable is the miscible flow. In the direct numerical simulations, disturbance growth occurs at much earlier times in the miscible flow, and also the miscible flow breaks spanwise symmetry more readily to go into three-dimensionality.
The following observations hold for both miscible and immiscible flows without surface tension. The stability of the flow is moderately sensitive to the location of the interface between the two fluids. The response is non-monotonic, with the least stable location of the layer being mid-way between the wall and the centreline. As expected, flow at higher Reynolds numbers is more unstable. 
\end{abstract}

\maketitle
\section{Introduction}
\label{sec:intro}

Two fluid flows display interesting instabilities due to viscosity and density contrasts between the fluids. Differences in these properties across the flow often exist simultaneously, but our interest is in isolating the effects of viscosity contrasts alone. The reverse case, of contrasting density but constant viscosity has been far more widely studied in geophysical and other contexts. Instabilities due to viscosity variation too have been investigated by several authors (see e.g.  \cite{yih67a,preziosi89a,chen96a,petitjeans96a,joseph97a,lajeunesse99a,bala05a,dolce08a,talon13a,john13a,valluri14a} for both immiscible and miscible  fluids. An extensive discussion of instability associated with such flows can be found in a recent review by \cite{rg2014}.

By conducting a linear-stability analysis \cite{yih67a} was the first to demonstrate that immiscible fluid layers (with a sharp interface between the two fluids) in shear flow are unstable to infinitesimally small long-wave disturbances at any Reynolds number. Since then instability in the context of sharp interfaces has been investigated by many researchers (e.g., \cite{hooper85a,hooper83a,valluri10a}), and short wave instabilities were found as well. The mechanism of this short wave instability was provided by \cite{hinch84a}.

Miscible flows are different from immiscible flows in an important way. The diffusivity, characterised by the inverse of the Schmidt number, is among the factors that plays an important role. \cite{govindarajan04a} investigated three-layer Poiseuille channel flows (wherein two miscible fluids are separated by a mixed region) and showed that at higher Schmidt numbers these flows go unstable at lower Reynolds numbers. She  found that when the more (less) viscous fluid occupies the near-wall regions, the flow is significantly destabilised (stabilised). These effects are accentuated by an increase in viscosity contrast. For pipe flow when the viscosity ratio is large, \cite{selvam07a} found that the flow can be destabilised even in the opposite scenario, i.e. when the less viscous fluid is near the wall. {\cite{ern03a} studied the influence of diffusion and mixed layer thickness in a miscible two-fluid Couette flow at a Reynolds number less than one, and showed that diffusivity has a nonmonotonic effect on the growth rate of the disturbance. They reported that flows at intermediate Schmidt numbers can be more unstable than flows at either very low or very high Schmidt numbers. They also found regimes at low Reynolds number where miscible flows are more unstable than those interfacial flow (with no diffusion across the sharp interface).} The geometry and flow considered in the present work is the same as that of \cite{talon11a}. They found that in the limit $\Re \to 0$, instability can be triggered by four different types of modes depending on the interface location. A mechanism of these instabilities in the Stokes flow regime was also provided based on perturbation of the concentration relative to the interface. The authors distinguish this instability from the inertial mechanisms of both \cite{hinch84a} and \cite{govindarajan04a}. \cite{talon11a} also remarked that the instability observed in their study is similar to the one of \cite{ern03a} in Couette flow. Instability at high Reynolds number of this miscible flow and its dynamics in the nonlinear regime has not been investigated yet. In the present study, we also investigate the difference between interfacial instabilities (without surface tension) with a viscosity jump across the interface, and instabilities at high Schmidt number (poor diffusivity) and  at high Reynolds number in pressure-driven two-layer miscible channel flow, by performing linear stability analyses and direct numerical simulations.

Important in our discussion will be the location of the critical layer (the layer at which the phase speed of the disturbance is close to  the mean streamwise velocity, and a major portion of the kinetic energy production takes place). In miscible flow, when this layer overlaps the viscosity-stratified layer, the dominant balance at the lowest order is changed,  \citep{govindarajan04a} so an additional `overlap' mode of instability can occur. When the two layers are well-separated, the Tollmien-Schlichting mode of instability is the most likely. In immiscble flows too, both scenarios can occur, i.e., the interface may or may not coincide with the critical location of the dominant disturbance.

The rest of the paper is organized as follows: The mathematical formulation of the linear stability equations for miscible and immiscible flows are presented in sections \ref{sec:formulation} and {\ref{sec:formulationi}, respectively. The budget for disturbance kinetic energy is formulated in section \ref{sec:energy}. The results are discussed in section \ref{sec:results}, and concluding remarks are given in section \ref{sec:conclusion}.

\section{Formulation: two miscible fluids with a mixed layer in between}
\label{sec:formulation}

\begin{figure}
\centering
 (a)  \\
\includegraphics[width=0.45\textwidth]{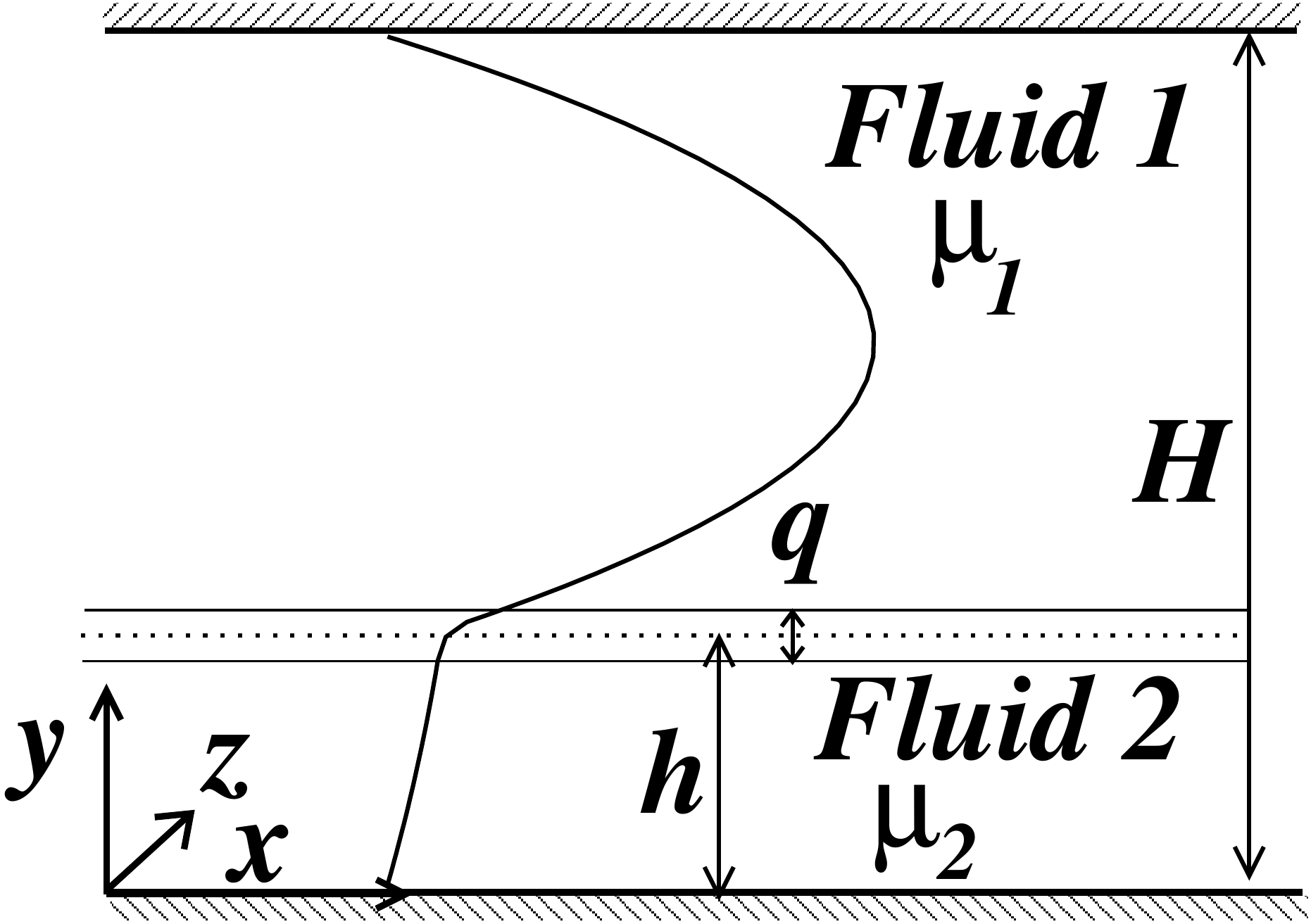} \\
 (b) \\
\includegraphics[width=0.45\textwidth]{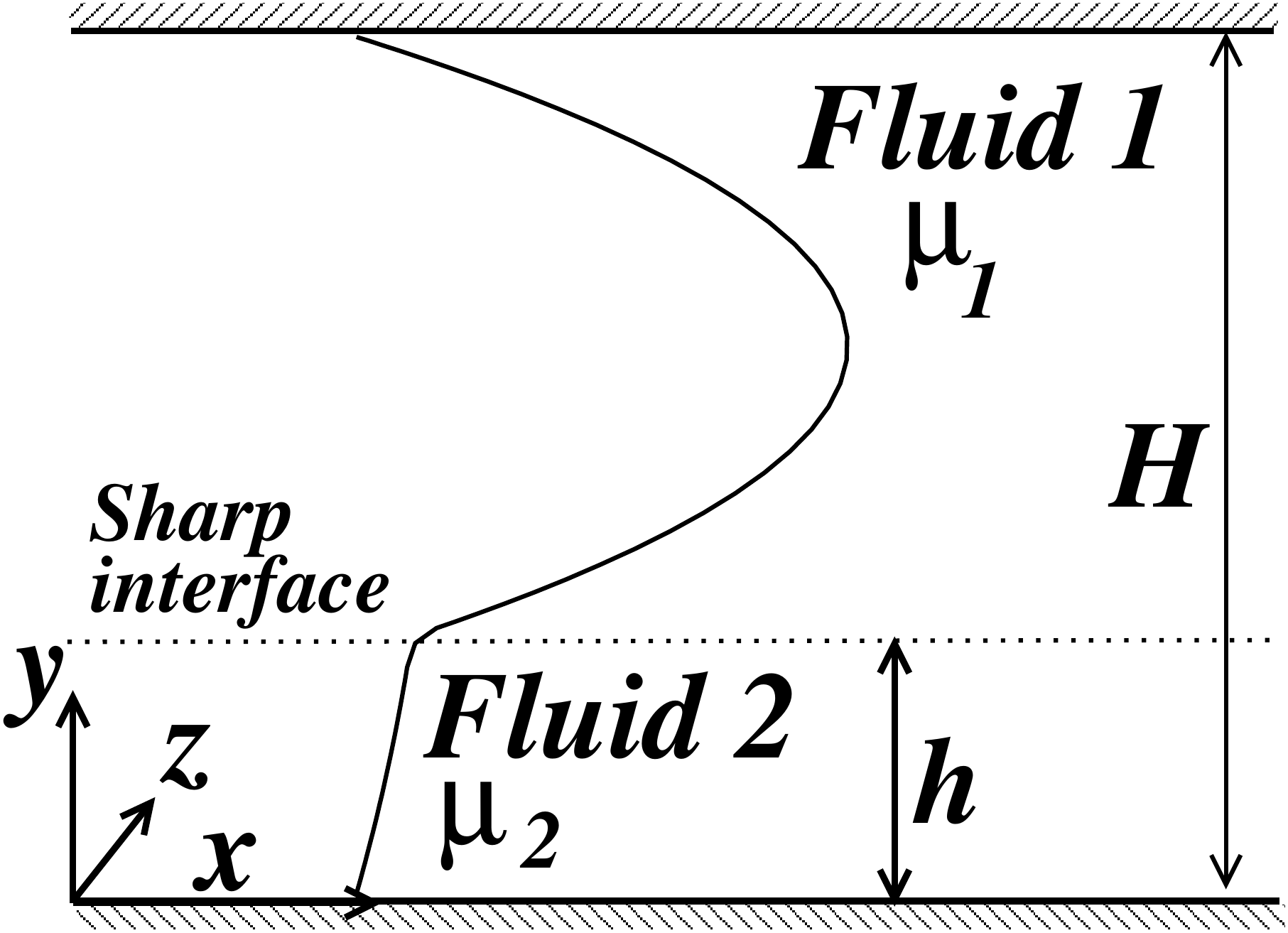} \\
\caption{Schematic of the flows. (a) Miscible case: Fluids `1' and `2' occupy the bottom ($0 \le y \le h-q/2$) and top ($h+q/2 \le y \le H$) layers, respectively. The two fluids are separated by a mixed layer of uniform thickness $q$. (b) The corresponding immiscible case.}\label{schematic}
\end{figure}  

The linear stability analysis and direct numerical simulation of a two-layer channel flow made up of two miscible, Newtonian and incompressible fluids of equal density and different viscosities is considered. The Cartesian coordinate system $(x,y,z)$ is used to formulate the problem, where $x$, $y$ and $z$ denote the coordinates in the horizontal, the vertical and spanwise directions, respectively. As shown in Fig. \ref{schematic}(a), the top (fluid `1' of dynamic viscosity $\mu_1$) and bottom (fluid `2' of dynamic viscosity $\mu_2$) fluids occupy the regions $0 \le y \le h-q/2$ and $h+q/2 \le y \le H$, respectively, where $q$ is the mixed layer thickness. The channel walls are located at $y=0$ and $y=H$, and periodic boundary conditions are imposed in the spanwise direction. The viscosity variation occurs due to the spatially varying magnitude of a scalar ($s$), which could be, for example, the concentration of a solute, or temperature. Without loss of generality, the base state concentration $s_0$ is taken to be 1 in the bottom layer, and 0 in the top layer, and it varies from 1 to 0 in the mixed layer. Thus, $q=H$ and $0$ represent complete stratification and a sharp interface, respectively. The latter is shown in Fig. \ref{schematic}(b). 

The viscosity, $\mu$, is modeled as an exponential function of the scalar $s$:
\begin{equation}
\mu= \mu_1 \exp {\left ( s R_s \right)},
\end{equation}
where $R_s \left(\equiv \ln \left ({\mu_2/\mu_1}\right)\right)$ is the log-mobility ratio of the scalar. The following scaling is employed to render the governing equations dimensionless:
\begin{eqnarray}
(x,y,z,q,h) = H \left({\widetilde x, \widetilde y, \widetilde z, \widetilde q, \widetilde h}\right),
\hspace{1mm} t={H^2 \over Q} \widetilde t, \nonumber \\
\hspace{1mm} (u,v) = {Q \over H} (\widetilde u, \widetilde v),
\hspace{1mm} p= {\rho Q^2 \over H^2} \widetilde p,
\hspace{1mm} \mu = \widetilde \mu \mu_{1},
\label{eq:scaling}
\end{eqnarray}
where the tildes designate dimensionless quantities; $Q$ denotes the total volume flow rate per unit distance in the spanwise direction; $u$, $v$ and $w$ are the velocity components in the $x$, $y$ and $z$ directions, respectively; $p$ denotes pressure; $\rho$ is the constant density and $t$ is time. The dimensionless governing equations (after dropping the tildes) are given by 
\begin{equation}
\nabla \cdot \u = 0, \label{NS1}
\end{equation}
\begin{equation}
\left [ {\partial \u \over \partial t} + \u \cdot \nabla \u \right] = -\nabla p + {1 \over \Re} \nabla \cdot \left [\mu (\nabla \u + \nabla \u^T) \right], \label{NS2}
\end{equation}
\begin{equation}
{\partial s \over \partial t} + \u \cdot \nabla s = {1 \over \Sc \Re} \nabla^2  s, \label{NS3}
\end{equation}
where  $\u$ is the velocity vector, ${\Re} (\equiv \rho Q /\mu_1)$ and ${\Sc} (\equiv \mu_1  / \rho {\cal D})$ are the Reynolds number and Schmidt number, respectively, wherein $\cal D$ is the diffusion coefficient of the scalar.

\subsection{Base state}
\label{subsec:base}
The base state corresponds to a steady, parallel, fully-developed flow, i.e. $U=U(y), V=W=0$, and $P$ is linear in $x$. Here, the base state quantities are designated by upper-case letters for the flow variables, and by the subscript $0$ for viscosity and $s$. In order to make the concentration of the scalar continuous up to the second derivative at $y=h-q/2$ and $y=h+q/2$, the mean scalar $s_0(y)$ is chosen to be fifth order polynomials in the mixed layer \citep{malik05a}: 
\begin{eqnarray}
s_0&=& 1, \qquad 0 \leq y \leq h-q/2, \nonumber \\
s_0&=&
\sum^{6}_{i=1}a_i y^{i-1}, \quad h-q/2 \leq y \leq h+q/2, \nonumber \\
s_0&=& 0, \qquad h+q/2 \leq y \leq 1,
\label{eq:C}
\end{eqnarray}
where the $a_i$'s $~(i=1,6)$ are given by
$$
a_1 = {(h+q/2)^3 \over q^5}  \left [ 6 (h-q/2)^2-3 (h-q/2) q+q^2 \right ], $$ 
$$ a_2 =  -{30 (h-q/2)^2 (h+q/2)^2 \over q^5}, 
$$
$$
a_3 = {60 (h-q/2) \over q ^5} (h+q/2) h,$$
$$
a_4 = -{10 \over q^5} \left [6 (h-q/2)^2 + 6 (h-q/2) q + q^2 \right ], 
$$
\begin{equation}
\quad a_5 = {30 \over q^5} h \quad {\rm and} \quad a_6 = -{6 \over q^5}. \label{eq:a16}
\end{equation}
We have confirmed that results indistinguishable from the present are obtained by using any other sufficiently smooth profiles, such as the error function or the hyperbolic tangent. For the parameter range of the present study (high Reynolds numbers and high Peclet numbers) the mixed layer diffuses very slowly, with a divergence angle of the order of $Pe^{-1}$. Thus the assumption of locally parallel flow and the ensuing use of a constant thickness mixed layer are extremely reasonable in this context, with errors of $O[Pe^{-1}]$. A brief description on the validity of parallel flow assumption is provided in the Appendix. The assumption will also be justified later in the form of comparisons with direct numerical simulations, where no such assumption is made.

The base state streamwise velocity profile $U(y)$ is obtained by solving the steady, fully-developed version of Eq. (\ref{NS2}) using no-slip and no-flux conditions at the wall and the centerline of the channel, respectively, i.e.,
\begin{equation}
{\Re} \left ( {dP \over dx} \right) =\left(\mu_0 U^\prime\right)^\prime,
\label{eq:steady}
\end{equation}
where  $\mu_0= e^{\left(R_s s_0\right)}$ and the prime represents differentiation with respect to $y$. The nondimensional pressure gradient $dP/dx$ is fixed by using $\int^1_0 U dy=1$.

\begin{figure}
\centering
(a)\\
\includegraphics[width=0.4\textwidth]{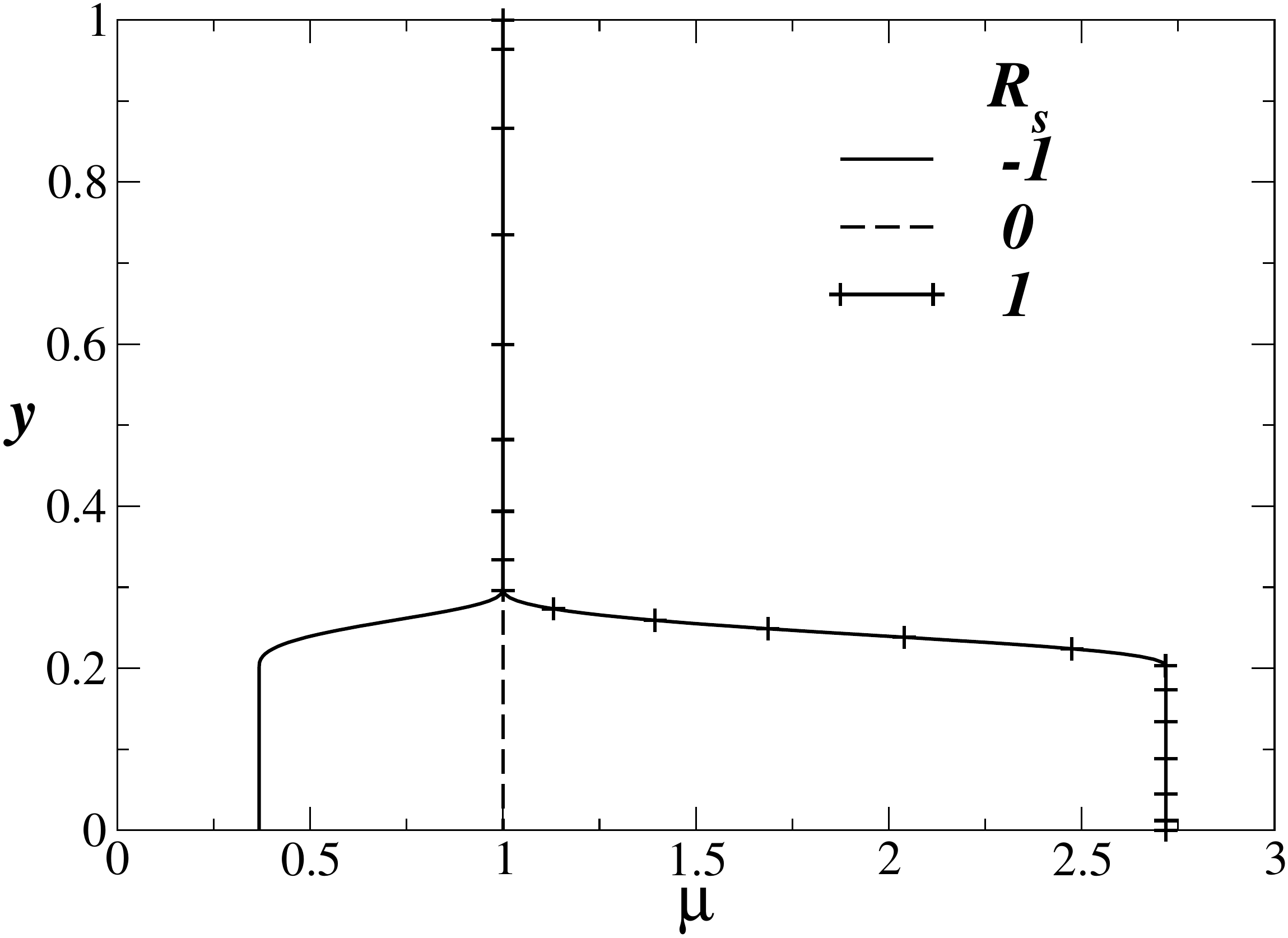} \\
(b) \\
\includegraphics[width=0.4\textwidth]{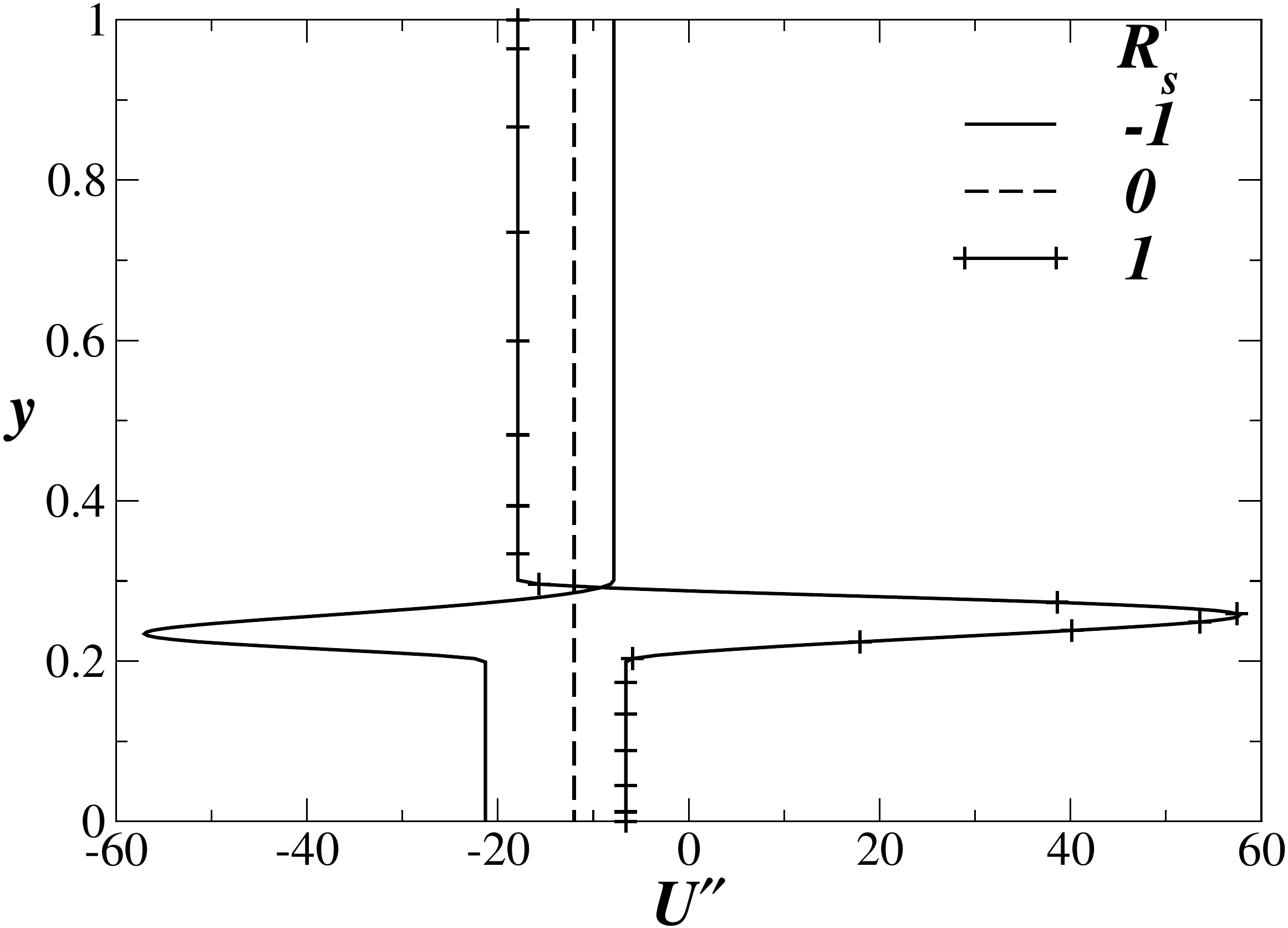}
\caption{Typical base steady-state profiles of (a) $\mu_0$ and (b) $U^{\prime\prime}$, for different values of $R_s$. The other parameters are chosen as $h=0.25$ and $q=0.1$.} 
\label{profile}
\end{figure}
In Figs. \ref{profile}(a) and (b) we show, respectively, typical profiles of the base state viscosity and of the second derivative of the mean velocity for different values of $R_s$. We choose to show the second derivative rather that the velocity profiles itself because it demonstrates that the velocity profile changes slope very rapidly in the mixed layer. Moreover, the case of 
$R_s=1$ is seen to contain a point of inflexion, which indicates a tendency for inviscid instability. 

\subsection{Linear stability analysis}
\label{subsec:linear}
The temporal linear stability of the base flow given by Eqs. (\ref{eq:C})-(\ref{eq:steady}) using a normal modes analysis considering two-dimensional perturbations is investigated. For a single fluid flow, Squire's theorem \citep{squire33}, states that every unstable three-dimensional disturbance is associated with an equally unstable two-dimensional disturbance at a lower value of the Reynolds number. We assume that in our stratified flow too, two dimensional disturbances go unstable at a lower Reynolds numbers than three-dimensional ones. We therefore study the linear instability to two-dimensional perturbations. Our direct numerical simulations confirm that in the range we study, two-dimensional disturbances are the first to go unstable. The flow variables are split into base state quantities and two-dimensional perturbations (designated by a hat):
\begin{equation}
(u,v,p,s)(x,y,t) = \left (U(y),0,P,s_0(y)\right ) + (\hat u, \hat v, \hat p, \hat s)(y) e^{\i\left({\alpha x} - \omega t \right)},
\label{split}
\end{equation}
and the perturbation viscosity is given by
\begin{equation}
{\hat\mu}= {d \mu_0 \over d s_0} {\hat s},
\end{equation}
where $\i \equiv \sqrt{-1}$, $\alpha$ and $\omega (\equiv \alpha c)$ are the wavenumber and frequency of the disturbance, respectively, wherein $c$ is the phase speed of the disturbance. In temporal stability analysis, $\alpha$ and $\omega$ are treated as real and complex quantities, respectively, whereas both are complex in spatio-temporal analysis (e.g. see for instance \cite{schmid01a}). We conduct the former here, where a given mode is unstable if $\omega_i > 0$, stable if $\omega_i < 0$ and neutrally stable if $\omega_i = 0$; $\omega_i$ being the imaginary part of $\omega$.

Following a standard approach by substituting Eq. (\ref{split}) into Eqs. (\ref{NS1})-(\ref{NS3}), subtracting of the base state equations, subsequently linearising and eliminating the pressure perturbation, we obtain the following linear stability equations \citep{govindarajan04a}, with the hat notation suppressed:
\begin{eqnarray}
\i \alpha \Re \left[ \left (\psi^{\prime\prime} - \alpha^2 \psi \right) \left(U- c\right) - U ^{\prime\prime} \psi \right ] = \nonumber
\\ \mu_0 \left ( \psi^{iv} - 2 \alpha^2 \psi^{\prime\prime} + \alpha^4 \psi \right ) +
\nonumber \\
2 {\mu_0^\prime} \left ( \psi^{\prime\prime\prime} -\alpha^2 \psi^\prime \right )+ \mu_0^{\prime\prime} \left ( \psi^{\prime\prime} + \alpha^2 \psi \right ) \nonumber \\ 
+ U^\prime \left ( \mu^{\prime\prime} + \alpha^2 \mu \right ) + 2 U^{\prime\prime} \mu^{\prime} + U^{\prime\prime\prime} \mu,
\label{stab1}
\end{eqnarray}
\begin{equation}
\i \alpha \Sc \Re \left [ \left(U - c \right ) s - \psi {s_0}^\prime \right ] = \left (s^{\prime\prime} - \alpha^2 s \right ),
\label{stab2}
\end{equation}
wherein the amplitude of the velocity disturbances are re-expressed in terms of a streamfunction [$(\hat{u}, \hat{v})=(\psi^{\prime},  -\i\alpha\psi)$]. 

{Solutions of these equations are obtained subject to the following boundary conditions at both the walls
\begin{eqnarray}
\psi&=&\psi^\prime=s^\prime = 0 \hspace{2mm} {\rm at } \hspace {2mm} y\pm1.
\label{bc1}
\end{eqnarray}
Eqs. (\ref{stab1})-(\ref{stab2}) along with the boundary conditions (\ref{bc1})) constitute an eigenvalue problem, which is solved using the public domain software, LAPACK.} A Chebyshev spectral collocation is used to discretised the domain. Due to the presence of large gradients in the viscosity-stratified region, a large number of grid points are required in this region. For this we use the stretching function proposed by \cite{govindarajan04a}:
\begin{equation}
y_j = {a \over \sinh (by_0)}\left[ \sinh \left\{(y_c-y_0)b \right\} + \sinh (b y_0) \right],
\label{stre}
\end{equation}
where $y_j$ are the locations of the grid points, $a$ is the mid-point of the stratified layer, $y_c$ is a Chebyshev collocation point,
\begin{equation}
y_0= {0.5 \over b} \ln \left [{1+(e^b-1)a \over 1+(e^{-b}-1)a} \right],
\end{equation}
and $b$ is the degree of clustering; $b=8$ is taken in this present study. The above formulation gives an accuracy of at least five decimal places in the range of parameters used.

\section{Formulation: two immiscible fluids separated by a sharp interface}
\label{sec:formulationi}

\subsection{Base state}
For pressure-driven flow of two immiscible fluids separated by a sharp interface (shown in Fig. \ref{schematic}(b)), the base state velocity profile is given by
\begin{equation}
U_1={1 \over 2} {{\left(\d P /\d x\right)^{-1}} \over e^{R_s}} \left[ {\d P \over \d x} y + c_3\right]^2+c_4,
\label{mean1}
\end{equation}
\begin{equation}
U_2={\d P \over \d x} {y^2 \over2}+c_1y+c_2, \label{mean2}
\end{equation}
with subscripts $1$ and $2$ denoting the lower and upper layers, respectively.
We obtained Eqs. (\ref{mean1}) and (\ref{mean2}) by integrating the steady, fully-developed dimensionless Navier-Stokes equations. Taking the undisturbed height of the interface to be $h_0$, the pressure gradient, ${\d P /\d x}$ and the integration constants, $c_1$, ${c_2}$, ${c_3}$, and ${c_4}$ are obtained by solving the following simultaneous equations, which correspond to no-slip conditions at the walls and balance of the tangential component of the stress at the interface.
$$
{{\left(\d P /\d x\right)^{-1}} \over {2 e^{R_s}}} \left \{\left[ {\d P \over \d x} h_0 + c_3 \right]^2-c_3^2 \right \}- $$ 
$${1 \over 2} {\d P \over \d x}\left({h_0}^2- 1\right)-c_1\left(h_0-1\right)=0,
$$
\begin{equation}
c_3=c_1, \quad c_2= -{1 \over 2}{\d P \over \d x}-c_1, \quad c_4=-{{\left(\d P / \d x\right)^{-1}} \over {2 e^{R_s}} c_1^2}. \label{eq:cequations}
\end{equation}
The pressure gradient, $dP/dx$, is obtained from the constant volumetric flow rate condition, i.e.,
\begin{equation} 
\int_0^{h_0} U_1 dy + \int_{h_0}^1 U_2 dy = 1.
\end{equation}

\subsection{Linear stability analysis}
We also examine the linear stability of the base state, obtained by solving Eqs. (\ref{mean1}) and (\ref{mean2}), to infinitesimal, two-dimensional disturbances. Each flow variable is expressed as the sum of a base state and a two-dimensional perturbation,
\begin{equation}
({\tilde u}_i, {\tilde v}_i, {\tilde P}_i)(x,y,t) = \left [U_i(y), 0, P_i \right] + \left (\hat u_i, \hat v_i, \hat p_i \right)(x,y,t),
\label{spliti}
\end{equation}
with $i=1,2$. Similarly the height $h$ of the interface can be expressed as,
\begin{equation}
h(x,y,t)=h_0+{\hat h} \quad {\rm with} \quad \hat{h}(x,t)=\tilde{h} e^{i(\alpha x - \omega t)}.
\label{split2}
\end{equation}
Substitution of Eqs. (\ref{spliti}), and (\ref{split2}) into the governing equations, and following the same procedure as before yields the following linear stability equations. In the lower layer:
$$\i \alpha {\Re} \left [\left \{ {v_1}^{\prime\prime} - \alpha^2 {v_1} \right \} (U_1-c)- U_1^{\prime\prime} {v_1}\right]= $$
\begin{equation}
e^{R_s} [ v_1^{\prime\prime\prime\prime}- 2 \alpha^2 v_1^{\prime\prime} +\alpha^4 v_1 ].
\label{stab1}
\end{equation}
In the upper layer:
\begin{equation}
\i \alpha {\Re} \left [\left \{ {v_2}^{\prime\prime} - \alpha^2 {v_2} \right \} (U_2-c)- U_2^{\prime\prime} {v_2}\right]=v_2^{\prime\prime\prime\prime}- 2 \alpha^2 v_2^{\prime\prime} +\alpha^4 v_2.
\label{stab3}
\end{equation}
The no-slip and no-penetration conditions at the walls can be written as
\begin{equation}
v_1 = {v_1}^\prime = 0 \quad {\rm at} \  y=0,
\label{nos}
\end{equation}
\begin{equation}
v_2 = {v_2}^\prime = 0 \quad {\rm at} \ y=1.
\end{equation}
The kinematic boundary condition gives
\begin{equation}
h = {{v_1} \over \i \alpha (U_1-c)} = {{v_2} \over \i \alpha (U_2-c)} \quad {\rm at} \ y=h.
\end{equation}
The continuity of the velocity components across the interface are expressed as
\begin{equation}
{v_1}^\prime - \i \alpha h {U_1}^\prime = {v_2}^\prime - \i \alpha h {U_2}^\prime \quad {\rm at} \ y=h,
\end{equation}
\begin{equation}
v_1 =v_2 \quad {\rm at} \ y=h.
\end{equation}
The normal stress jump and continuity of the tangential stress balance in the streamwise and spanwise directions are respectively given by
$$
\i \alpha {\Re} \left [\left \{ {v_1}^\prime (c-U_1)+ {U_1}^\prime {v_1} \right \}- \left \{ {v_2}^\prime (c-U_2)+{U_2}^\prime {v_2} \right \}\right] -$$
$$
2 {{\mu_1}} \alpha^2 {v_1}^\prime + 3 \alpha^2 {v_2}^\prime+e^{R_s} \left[{v_1}^{\prime\prime\prime}-\alpha^2{v_1}^\prime \right] -{v_2}^{\prime\prime\prime}=
$$
\begin{equation}
\alpha^4 \Gamma { ({v_2}^\prime -{v_1}^\prime) \over \i \alpha (U_2^\prime - U_1^\prime)} \quad {\rm at} \ y=h,
\end{equation}
$$
e^{R_s} \left [{v_1}^{\prime\prime}+\alpha^2{v_1} \right] - {(e^{R_s} {U_1}^{\prime\prime} - {U_2}^{\prime\prime}) \over (U_1-c)} {v_1}  
$$
\begin{equation}
={v_2}^{\prime\prime}+\alpha^2{v_2}\quad {\rm at} \ y=h.
\label{bcnew}
\end{equation}
Here $\Gamma \equiv {\gamma H / \mu_1 Q}$ is an inverse capillary number, in which $\gamma$ denotes the interfacial tension. {The complete derivation and linearisation of the stability equations can be found in \cite{sahu10a}.} In this work, we set $\Gamma$ to zero, because we wish to compare the miscible and immiscible cases without the additional factor of surface tension in the latter.

\section{Budget of disturbance kinetic energy}
\label{sec:energy}

An energy budget analysis can highlight the physical differences between the two flows in their stability behaviour.
A budget of disturbance kinetic energy, neglecting the surface-tension and gravity, is given by
$$
2 \omega_i  {1 \over \lambda} \int_a^b \int_0^\lambda E dx dy = {1 \over \lambda} \int_a^b \int_0^\lambda P dx dy +$$
\begin{equation}
 {1 \over \lambda \Re} \int_a^b \int_0^\lambda D dx dy+ I,
\label{energybudget}
\end{equation}
where $\lambda \equiv 2 \pi / \alpha$. For fluid 1 and fluid 2 ($a=h, b=1$) and ($a=0, b=h$), respectively. The kinetic energy, the rate of its production, and the rate of dissipation are given respectively by
\begin{equation}
E = {1 \over 2} \left (u^2 + v^2\right),
\end{equation}
\begin{equation}
P =   - u v {d U \over d y},
\end{equation}
and
\begin{equation}
D =  2 \mu \left [ \left ( {\partial u \over \partial x} \right)^2  +  \left ( {\partial v \over \partial y} \right)^2 + {1 \over 2}  \left ( {\partial u \over \partial y} +  {\partial v \over \partial x} \right)^2 \right ].
\end{equation}
The viscous work done by the mean flow on the interface is given by
\begin{equation}
I = {1 \over \lambda \Re} \int_0^\lambda \left [ u_1 \tau_1^{xy} - u_2 \tau_2^{xy} \right ]dx, \quad (at \hspace{2mm}  y= h),
\label{int}
\end{equation}
wherein $$\tau^{xy}= \mu \left ({\partial u \over \partial y} +{\partial v \over \partial x} \right).$$
Continuity of shear stresses implies that for the mean flow there is a jump in the slope of $U$, i.e.,
\begin{equation}
U_1^\prime\big|_{y=h} = \exp(R_s) U_2^\prime\big|_{y=h},
\end{equation}
and for the disturbance 
\begin{equation}
\tau_1^{xy}\big|_{y=h} = \tau_2\big|_{y=h}^{yx}=\tau^{xy}.
\end{equation}
Thus, Eq. (\ref{int}) can be written as
\begin{equation}
I = {1 \over \lambda \Re} \int_0^\lambda \tau^{xy} \left [ u_1 - u_2  \right ]dx, \quad ({\rm at} \hspace{2mm}  y= h),
\label{int2}
\end{equation}
When the interface is being deformed, streamwise disturbance velocities of unequal size are forced at the interface, i.e., $u_1 \ne u_2$, due to which energy transfer occurs from the mean flow to the disturbance. This quantity will remain positive even if the fluid layers are interchanged \citep{boomkamp96a}. 

For miscible flow, we would have the same expressions for $E$, $P$ and $D$, but integrated across the entire channel, and of course $I=0$.

Next we evaluate how miscible and immiscible two-fluid flows differ in their stability behaviour. We then perform direct numerical simulations and show that the nonlinear behaviour is consistent with the predictions of linear instability. The simulations also help us to estimate how three-dimensional the flow is.

\section{Results}
\label{sec:results}

\subsection{Linear stability analysis}
A log viscosity ratio of $R_s > 1$ gives rise to a velocity profile with a point of inflexion for $h<0.5$, as seen in Fig. \ref{profile}. Such a profile is likely to be more unstable than one without a point of inflexion, and therefore be the more interesting case, so we restrict ourselves to positive values of $R_s$. A typical set of disturbance growth rates is presented in Fig. \ref{effectRe}. The growth rates of the most unstable eigenmode are plotted as functions of wavenumber, for different values of Reynolds number. The instability behaviour for both miscible and immiscible two-fluid flows are shown in the same figure. As is usual in shear flows, the instability gets more severe as the Reynolds number increases. More remarkable is the fact that, for higher Reynolds numbers $(\Re \ge 200)$, the miscible flow is more unstable than the flow containing the immiscible interface.
\begin{figure}
\centering
\includegraphics[width=0.45\textwidth]{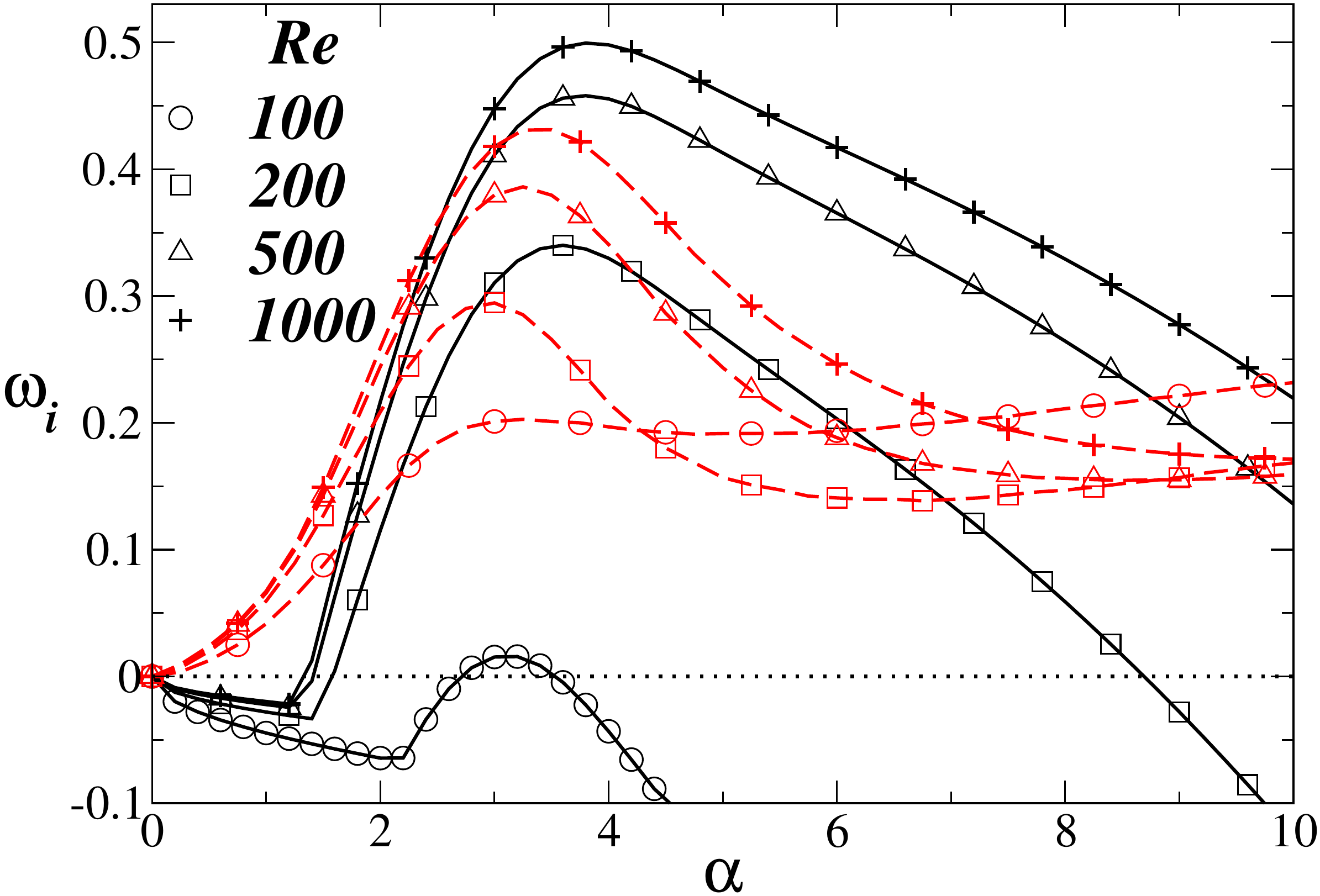} 
\caption{{Growth rates, $\omega_i$, of the most unstable disturbance as functions of the wavenumber, $\alpha$, for different values of the Reynolds number $\Re$ for $\Sc=10$, $q=0.02$, $R_s=1$ and $h=0.15$. The solid lines represent miscible flow, and red dotted lines represent the result for immiscible two fluid flow, with the interface placed at the same value of $h$, and with the same $R_s$. The symbols for a given Reynolds number are the same for miscible and immiscible flows.}}
\label{effectRe}
\end{figure}

\begin{figure}
\centering
(a)  \\
\includegraphics[width=0.45\textwidth]{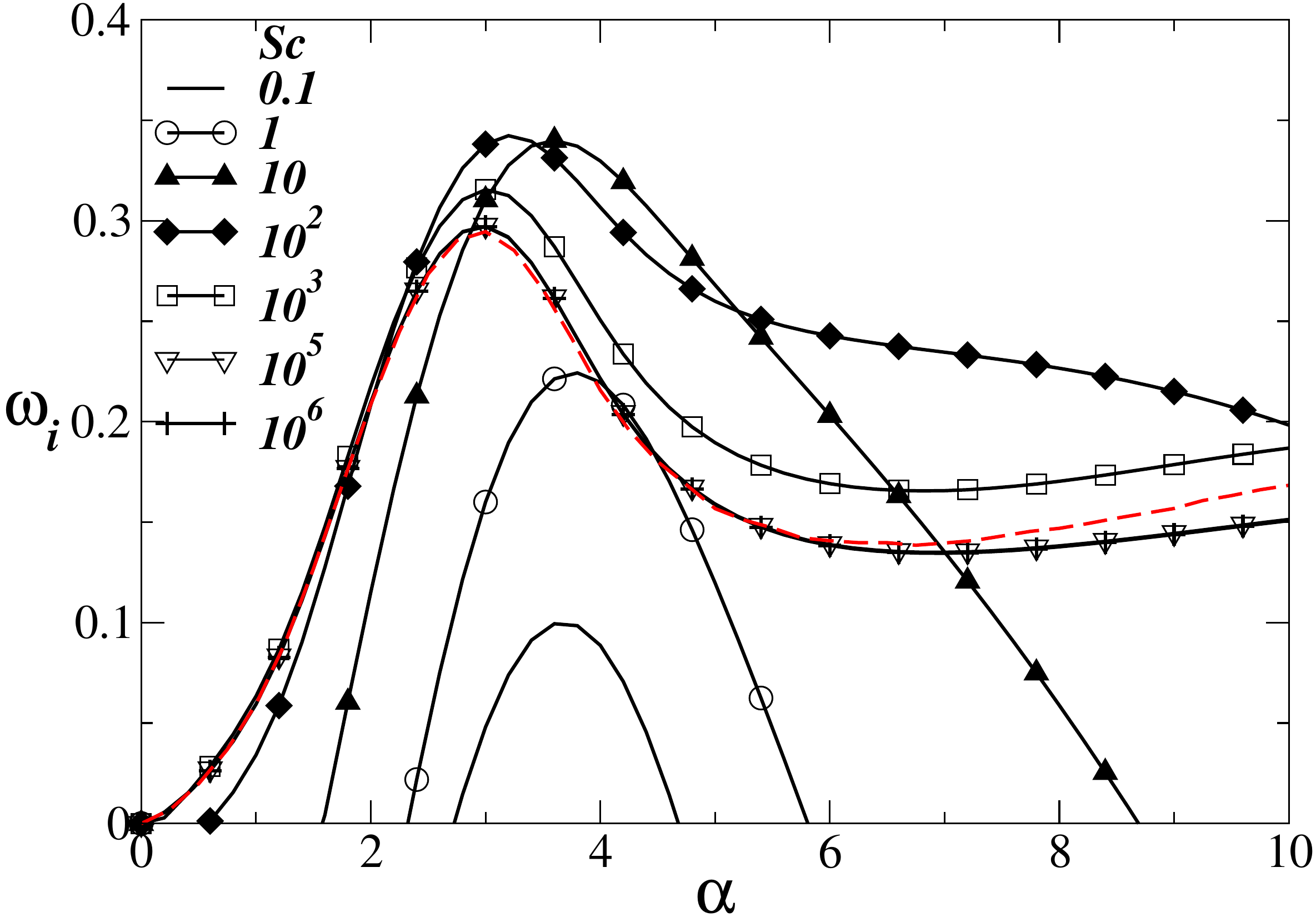} \\ 
 (b) \\
\includegraphics[width=0.45\textwidth]{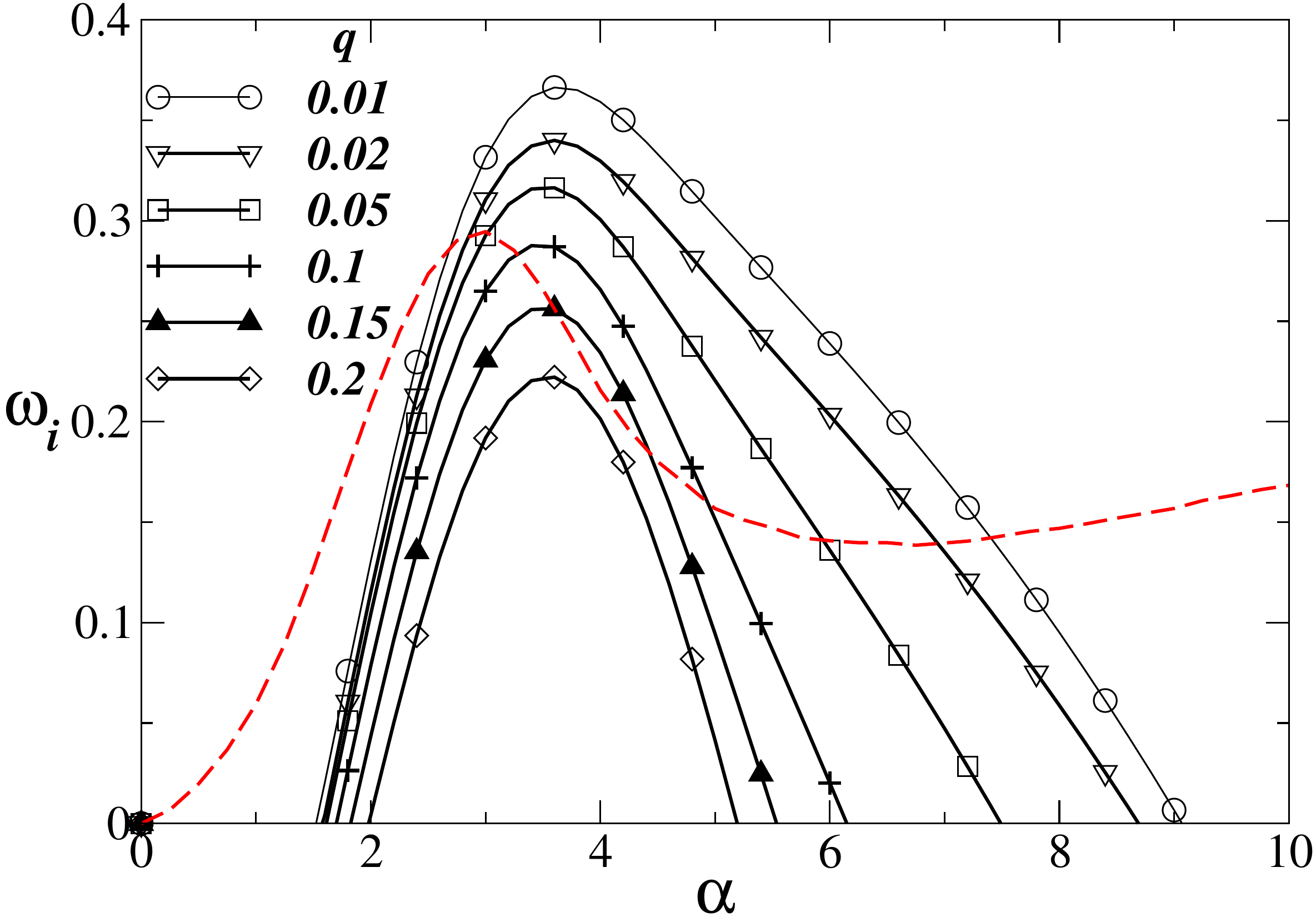}\\
\caption{The dispersion curves ($\omega_i$ versus $\alpha$) for different values of (a) $\Sc$ for $q=0.02$, and (b) $q$ for $\Sc=10$. The rest of the parameter values are $\Re=500$, $R_s=1$ and $h=0.15$. The dotted lines represent the results for the immiscible case.} \label{effectSc2}
\end{figure}

\begin{figure}
\centering
(a)  \\
\includegraphics[width=0.45\textwidth]{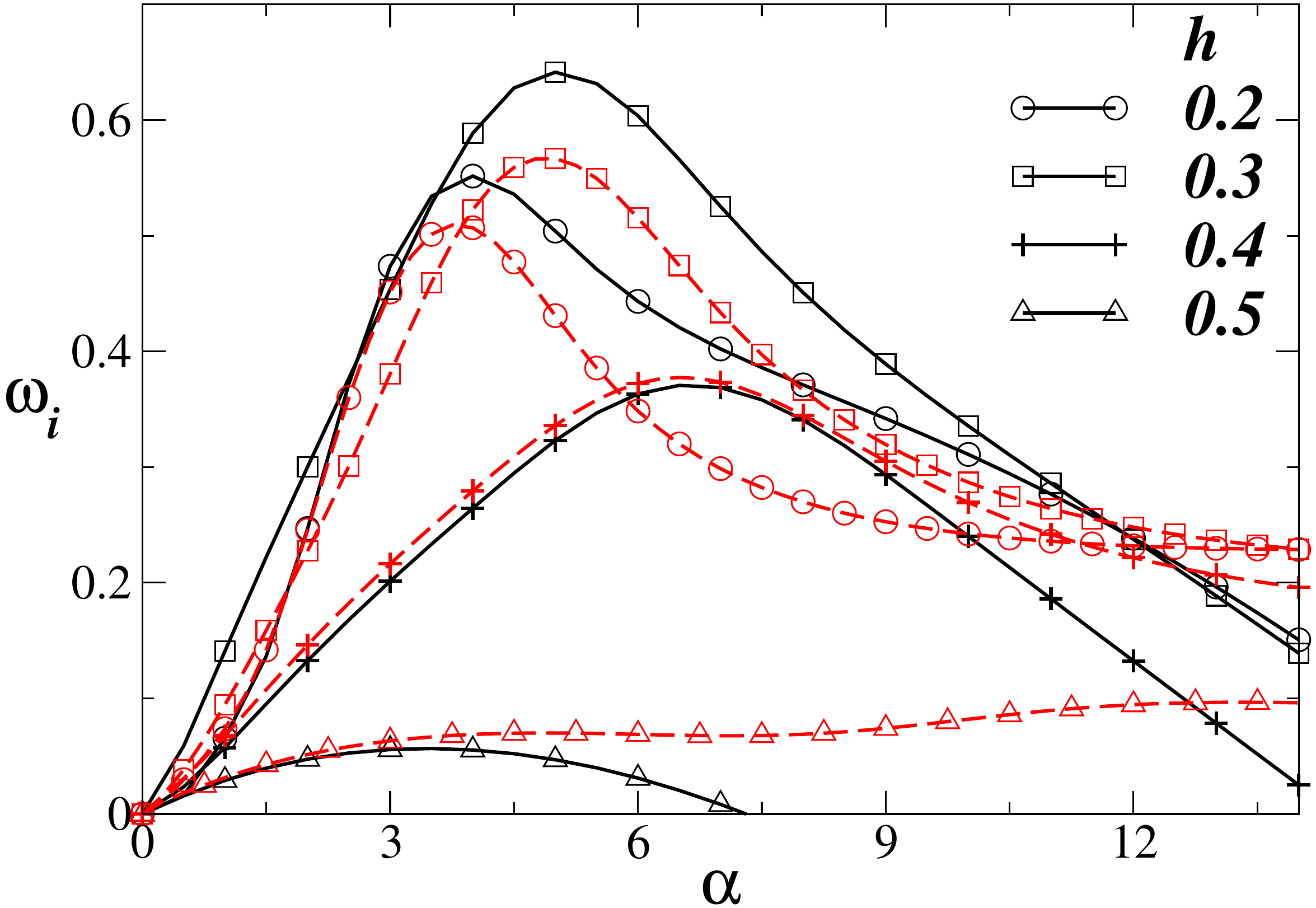} \\
(b) \\
\includegraphics[width=0.45\textwidth]{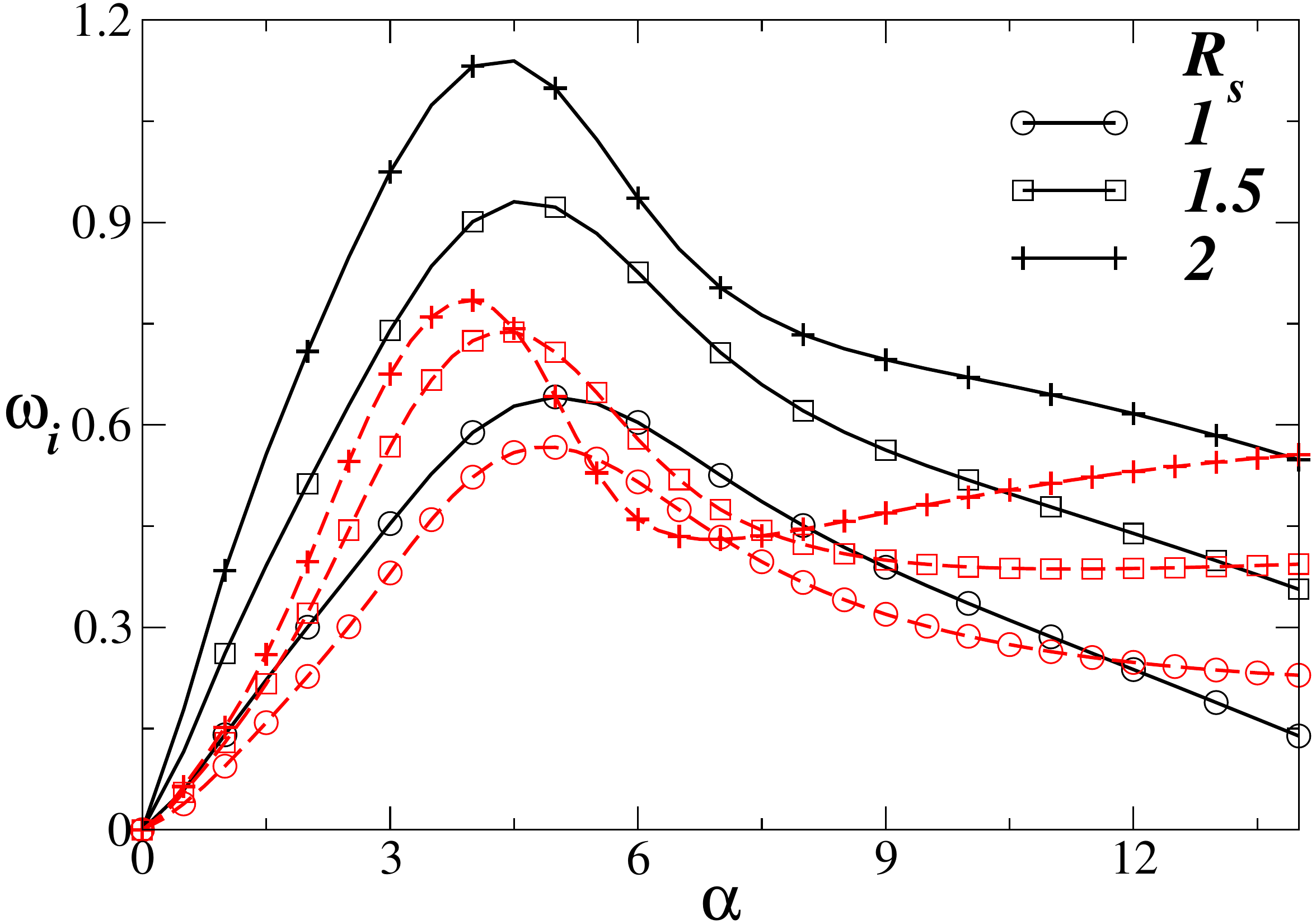}
\caption{Dispersion curves ($\omega_i$ versus $\alpha$) for different values of (a) $h$ for $R_s=1$, (b) $R_s$ for $h=0.3$. The rest of the parameter values are $\Re=500$, $\Sc=100$ and $q=0.01$. The red dotted lines show the corresponding growth rates in the case of immiscible flow.} \label{effecth}
\end{figure}

It is now accepted knowledge that shear flows of two or more fluids most often become more unstable at high Schmidt numbers (when the diffusivity of one fluid in another is very low). The expectation therefore would be that if we increase the Schmidt number of the miscible flow, flow would become increasingly unstable. We see in Fig. \ref{effectSc2}a, that the behaviour is not monotonic with increase in Schmidt number. While the flow becomes more unstable as we increase the Schmidt number up to a value of $100$, a further increase in $\Sc$ decreases the growth rate of the most-unstable mode. For very high $\Sc$, i.e., for $\Sc>10^5$, the behaviour of the most-unstable mode is the same as that of the immiscible flow. Thus the immiscible case is less unstable than two-fluid flow of intermediate miscibility. We will show later in this section that the overlap of the mixed layer with the critical layer is the underlying mechanism in the present system, which is characteristic of high Reynolds number flow. A non-monotonic response to change in Schmidt number was also obtained by \cite{ern03a} in Couette flow at low Reynolds number, but their mechanism was not that of the present, as will be discussed below.

Another parameter which is known to affect flow stability significantly is the thickness $q$ of the mixed layer. In Fig. \ref{effectSc2}(b), we see that as the mixed layer is made thinner, the growth rate of the dominant instability increases. This is as expected, and is caused by the fact that as $q$ decreases, the viscosity gradient becomes sharper, making the stability operator more singular. In this figure, $\Sc=10$ is used, as a typical example. It can be seen that the growth rate remains sensitive to $q$ at all values of $q$ that we have considered. Here too it can be observed that the dispersion curve for the immiscible flow (shown by the  dashed line) is well below the dispersion curves of the miscible system for $q \le 0.05$. This figure is for a Reynolds number of $500$, but we have repeated all our calculations at a Reynolds number of $1000$ as well (not shown), and the behaviour is qualitatively the same. Again, when the layer is thin enough, flow of intermediate miscibility is significantly more unstable than the immiscible case.

In Fig. \ref{effecth}(a), we investigate the effect of $h$, the height of the mixed layer from the bottom wall. When the interfacial layer is close to the bottom or top walls, the flow is stabler than when the mixed layer is near the middle, and a value of $h \sim 0.3$ is the least stable. The response to the location of the interfacial layer is thus non-monotonic. For $h<0.4$ we see that the miscible flow is more unstable than the immiscible. It can be seen in Fig. \ref{effecth}(b) that immiscible flow is not very sensitive to viscosity ratio, but disturbances in miscible flow grow much faster at higher viscosity ratios. For all the viscosity ratios considered, it is seen that the the miscible flow is more unstable than the corresponding immiscible flow. Taking into consideration all the linear stability results, we see that our finding that miscible flow (at intermediate levels of miscibility) is more unstable than the immiscible flow is a general result for high Reynolds number channel flow of two-fluids. 

In order to investigate the instability mechanism, neutral stability curves for different values of $\Sc$ and $h$ are plotted in Fig. \ref{neutralcurves}(a) and (b), respectively. The standard Tollmien-Schlichting mode contributes a region of instability, seen on the extreme right of the plots, i.e., at high Reynolds number. In addition, a distinct region of instability is observed, which grows in size with increase in Schmidt number (Fig. \ref{neutralcurves}(a)). The phase speed in this regime is close to the mean velocity in the mixed-fluid layer (we shall return to this point in Fig. \ref{overlap}). Note that the neighbourhood of thickness $O(R^{-1/3})$, where the phase speed of the dominant disturbance is close to the mean velocity, is the critical layer where most of the disturbance kinetic energy is produced. It was shown in \cite{govindarajan04a} for a three-layer channel flow that the above condition, of an overlap between the critical layer with the mixed layer, contributes to a singular perturbation term in the stability operator. The resulting new mode of instability was termed the ``overlap'' mode. The energy production is interfered with in a major way by this overlap. Since this instability is inherently inertial it is distinct from the modes obtained by \cite{talon11a} for Stokes flow. Besides the fact that Talon \& Meiburg found a similarity with their instability and that of \cite{ern03a}, we may check directly whether there is an overlap mechanism operational in the latter. It can be checked that the critical layer obtained by \cite{ern03a} is well below their mixed layer. Thus the instabilities observed by \cite{talon11a,ern03a} are not overlap modes. In fact in Stokes flow, the dissipation of the overlap mode would be infinite and energy production could never exceed dissipation in order to make the flow unstable.

 \begin{figure}
\centering
 (a)  \\
\includegraphics[width=0.45\textwidth]{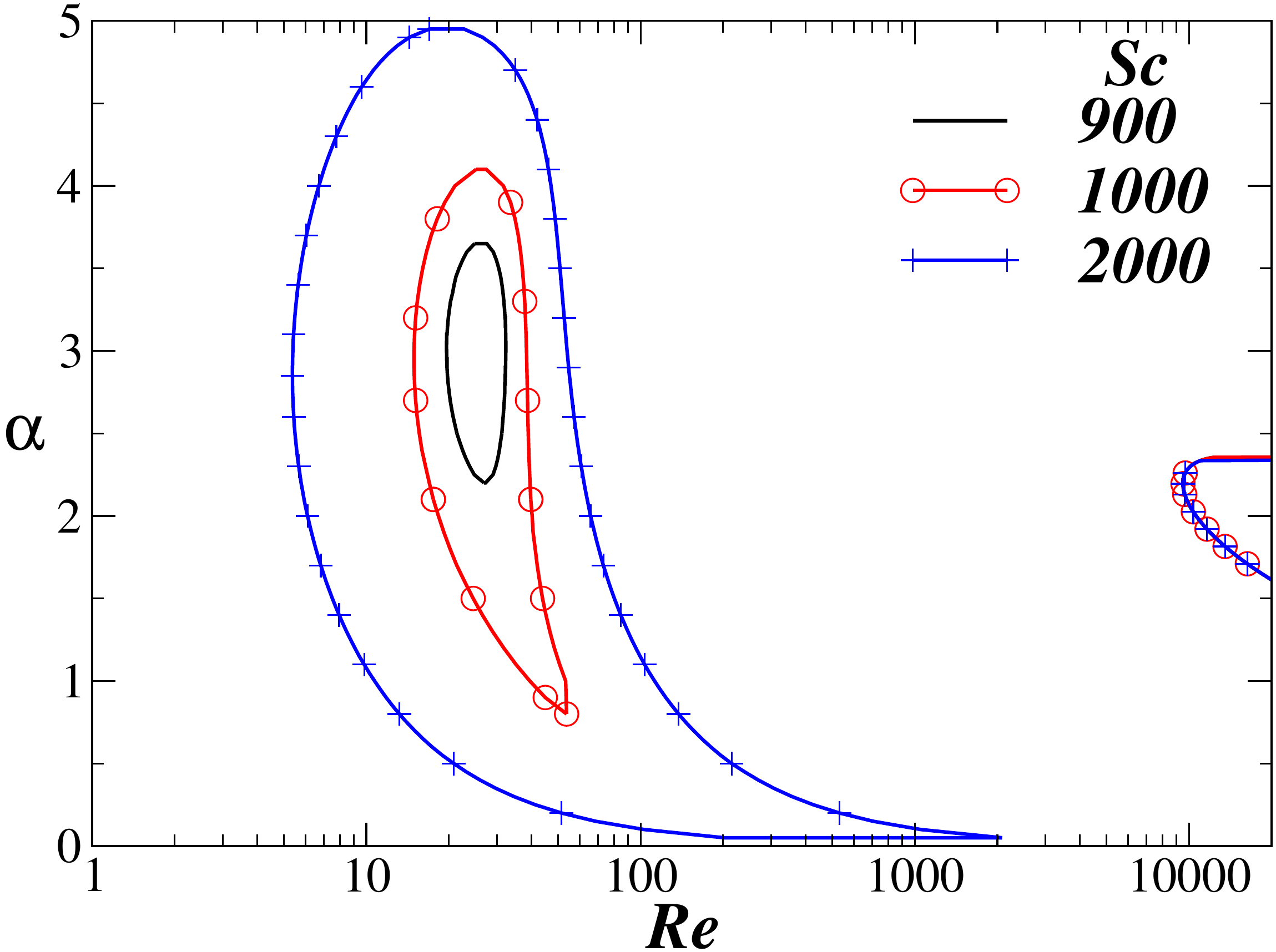} \\
 (b) \\
   \includegraphics[width=0.45\textwidth]{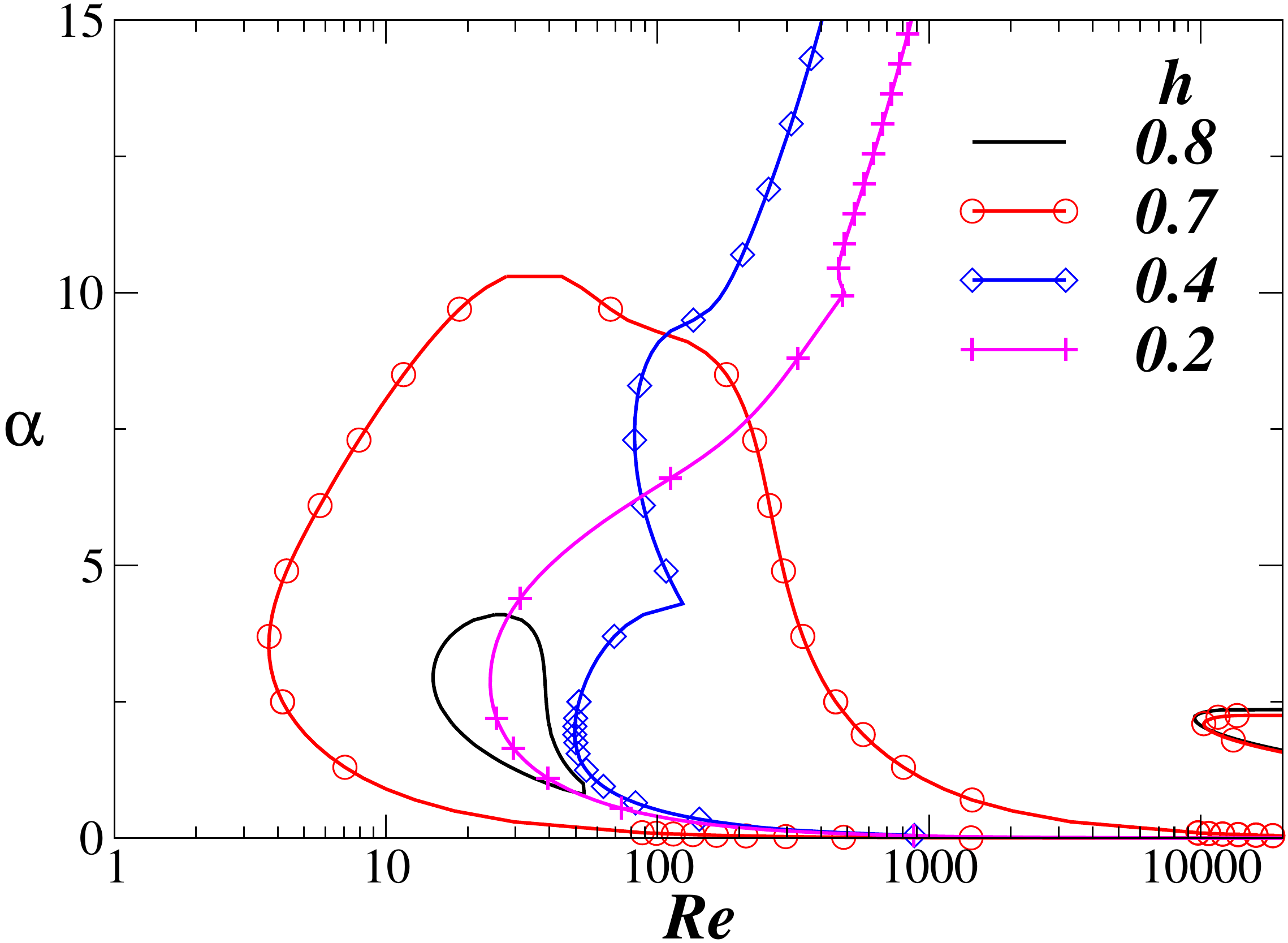}
\caption{Neutral stability curves for different values of (a) $\Sc$ for $h=0.8$, and (b) $h$ for $\Sc=1000$. The rest of the parameter values are $R_s=0.6$ and $q=0.05$. For the boundaries which appear as closed curves, the region contained within is unstable, and the outside is stable. For the open curves, the region to the right is unstable, and that to the left is stable. The curves at high Reynolds number (of about $10000$, where $\alpha \sim 2$) correspond to the Tollmien-Schlichting mode of instability, whereas the other curves correspond to the overlap mode, as defined in \cite{govindarajan04a}.} 
\label{neutralcurves}
\end{figure}

It is seen in Fig. \ref{neutralcurves}(b) that the overlap mode displays a distinct region of instability when $h \ge 0.7$, whereas for lower values of $h$ a much larger region is unstable, and it is difficult to distinguish the ``overlap" mode anymore, except that it can be recognised by an apparent kink  in the neutral boundary. This behaviour is observed over a range of parameters, and an example at low $\Sc$ is shown in Fig. \ref{overlap} (a) for $R_s=1$, $h=0.2$ and $q=0.05$. To verify whether some part of the neutral boundary corresponds to an overlap mode of instability, we examine Fig. \ref{overlap}(b), which represents behaviour along the lower limb of the neutral stability boundary of Fig. \ref{overlap}(a). The distance between the centres of the mixed layer and the critical layer is the quantity $h-y_{cr}$. This quantity is plotted versus Reynolds number along the lower limb of the neutral stability boundary in this figure. The width of the critical layer may be estimated as $\sim (U^\prime_{cr} \Re \alpha)^{-1/3}$, and this is denoted by the region within the red lines. It is now evident that different modes of instability are in operation on either side of the kink seen in the neutral stability boundary in Fig. \ref{overlap}(a). The neutral mode at low Reynolds numbers has a small distance between $y_{cr}$ and $h$. In fact this distance is seen to be smaller than the critical layer thickness over a range of Reynolds numbers, indicating that overlap effects must be in operation at the lowest order. At high Reynolds number however (beyond the kink, coming downwards along the neutral boundary), a sudden jump is seen in $h-y_{cr}$, and this difference is greater than the critical layer thickness, indicating that different physics is operational there.

 \begin{figure}
\centering
(a)  \\
\includegraphics[width=0.45\textwidth]{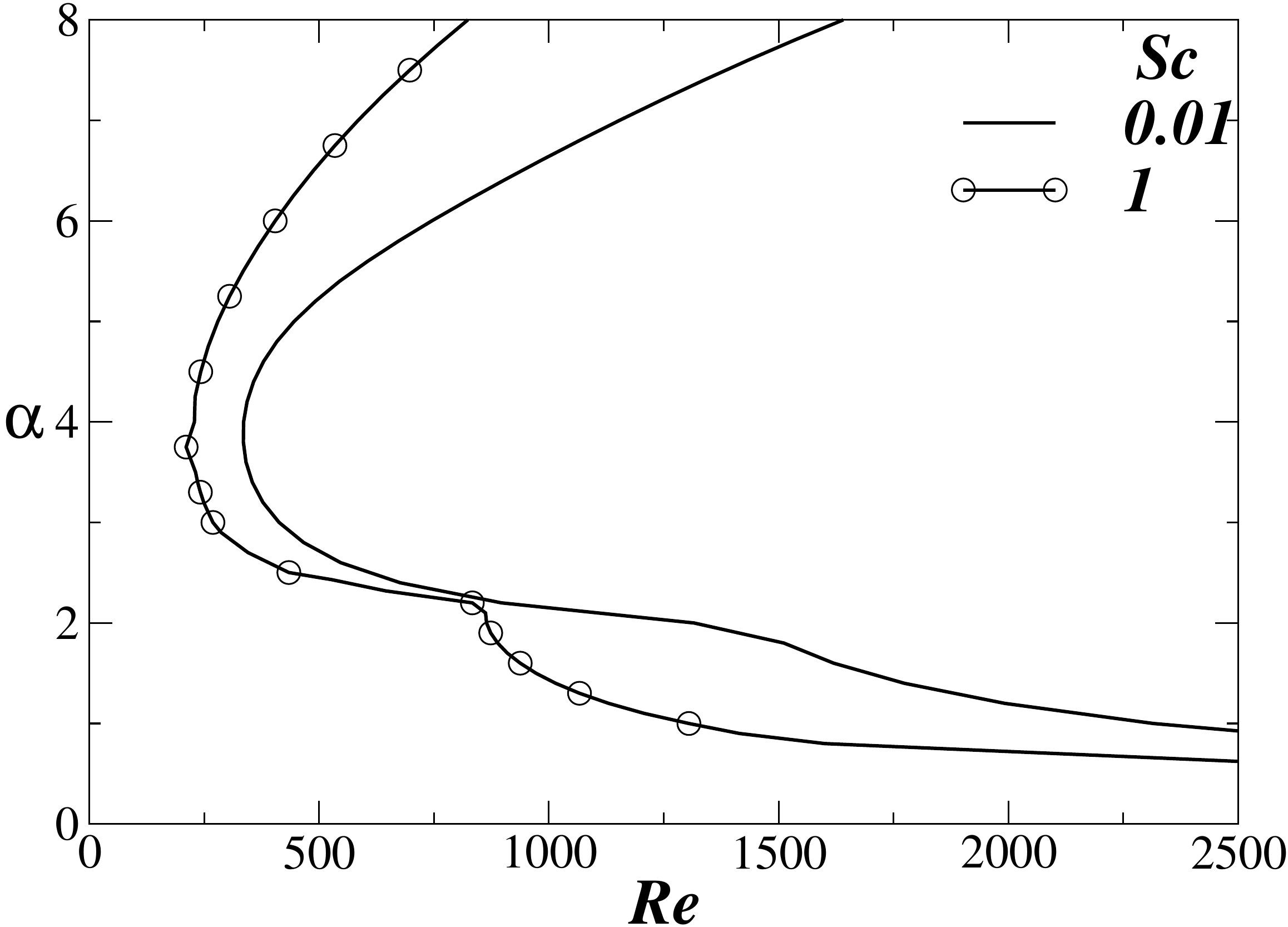} \\
 (b) \\
\includegraphics[width=0.45\textwidth]{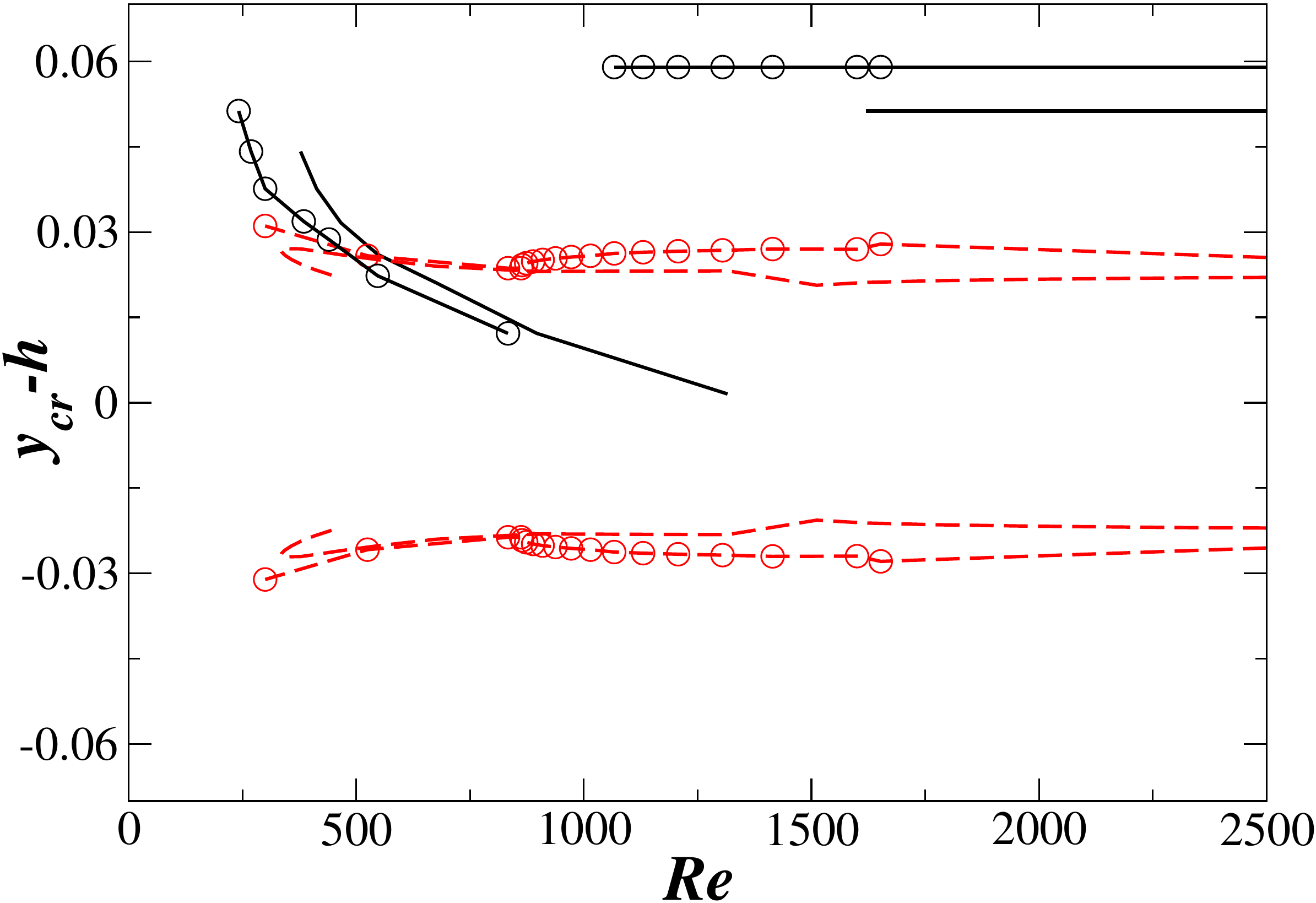} 
\caption{(a) Neutral stability curves, (b) $y_{cr} -h$ versus $\Re$ along the lower limb of the neutral stability curves (shown in panel (a)). Without symbols: $\Sc=0.01$, with symbols: $\Sc=1$. The rest of the parameter values are $R_s=1$, $h=0.2$ and $q=0.05$. The regions contained within the red lines correspond to an order of magnitude estimate of the critical layer thickness $\pm{1 \over 2} (U^\prime_{cr} \Re \alpha)^{-1/3}$. Here $y_{cr}$ is the location of the critical layer; i.e., the $y$ at which $U=c_r$.} \label{overlap}
\end{figure}

The production of disturbance kinetic energy is examined next for the intermediate Schmidt number case of Fig. \ref{effectRe}. The variations of the disturbance kinetic energy rate $E$, the production rate $P$ and the dissipation rate $D$ across the channel are displayed in Figs. \ref{energy2}(a), (b) and (c), respectively. At $R_s=1$, the maximum growth rate in both miscible and immiscible flows occurs at a wavenumber $\alpha \sim 4.5$, so the most dangerous mode at $\alpha=4.5$ is chosen to do this energy budget analysis. The kinetic energy production is seen in Fig. \ref{energy2}(c) to peak close to the location of the mixed layer, indicating that overlap effects are in operation. The difference between the production and the dissipation rates, which gives, in the miscible flow case, the change disturbance kinetic energy per unit time, is shown in Fig. \ref{energy2}(d). The production and the dissipation in the portions of the channel away from the interfacial or mixed layer are very similar in the two flows. The major difference is apparent in the vicinity of the interfacial or the viscosity stratified layer. It is clear that the net production $P-D$ in the miscible case exceeds that in the immiscible case in the mixed region. The immiscible flow has an additional contribution $I$ to disturbance growth. We denote by ${\cal E}$, ${\cal P}$ and ${\cal D}$ the integrals of $E$, $P$ and $D$ across the channel from wall to wall. It is seen from Eq. (\ref{energybudget}) that the growth rate is given by $2\omega_i =   ({\cal P} - {\cal D}+I)/{\cal E}$. $\cal P$, $\cal D$ and $I$ normalised with $\cal E$ for the immiscible case are 0.7479, -0.6054 and 1.03423, respectively. This gives the growth rate $\omega_{i,max}=0.5878$, where the subscript $max$ stands for the maximum growth rate of the fastest growing mode. For the miscible case: the values of $\cal P$ and $\cal D$ normalised with $\cal E$ are 5.6366 and -3.5407, respectively, giving $\omega_{i,max}=1.045$. It is thus seen that the net production minus dissipation on disturbance kinetic energy in the miscible case is larger than the contribution of all terms in the immiscible case.

\begin{figure}
\centering
 (a) \hspace{2.5cm} (b) \\
\includegraphics[width=0.22\textwidth]{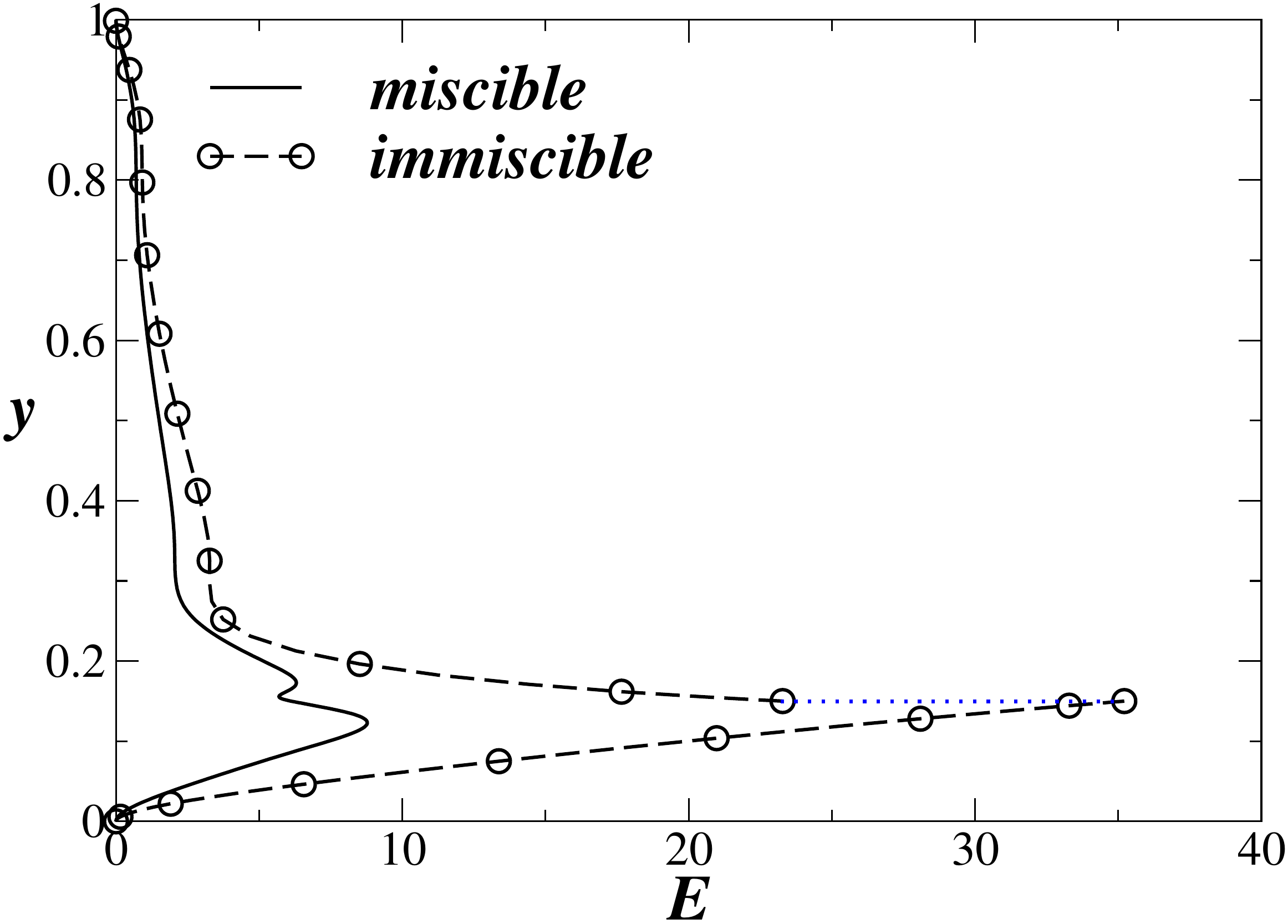} \hspace{2mm} \includegraphics[width=0.22\textwidth]{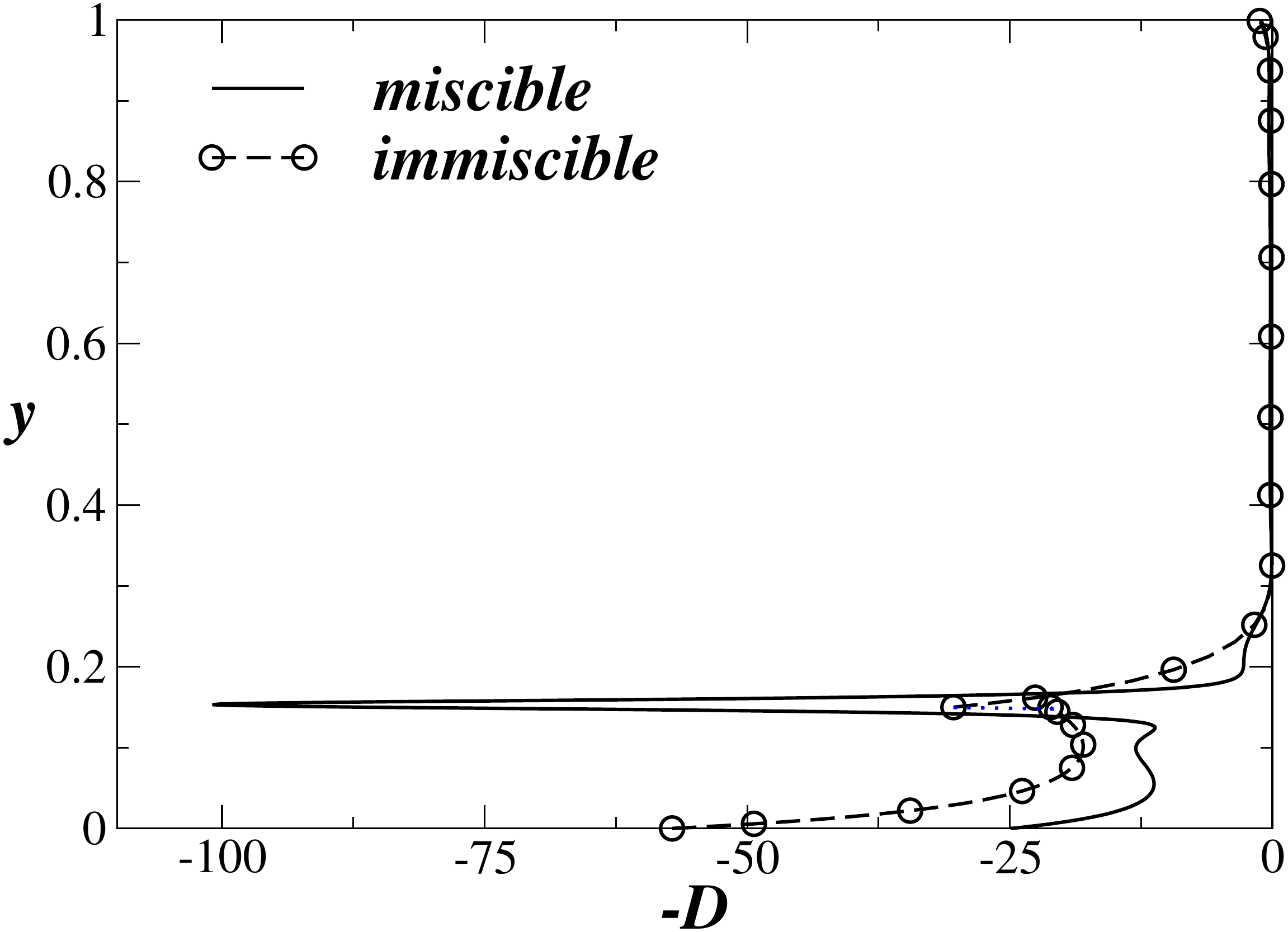} \\
 (c) \hspace{2.5cm} (d) \\
\includegraphics[width=0.22\textwidth]{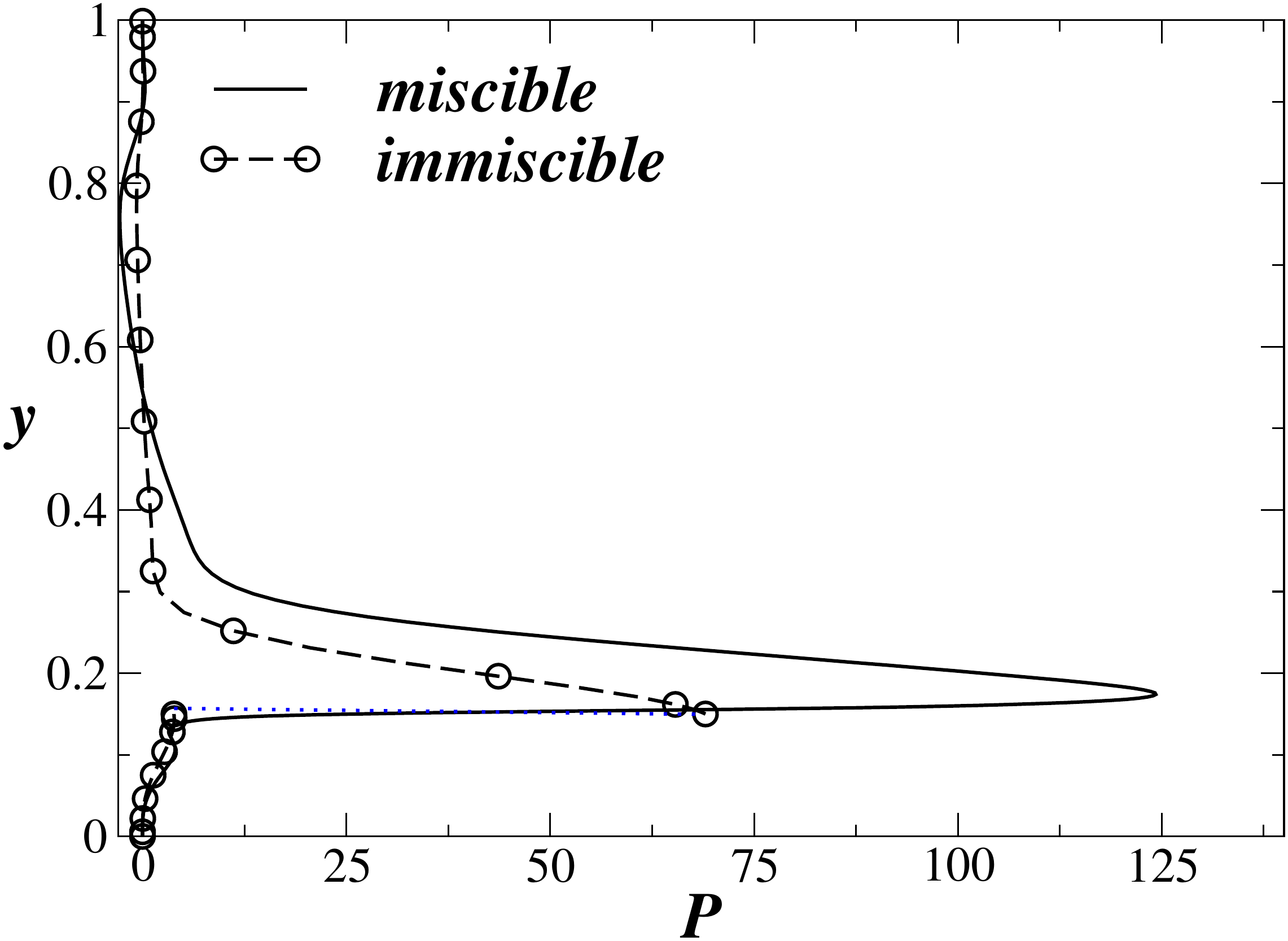} \hspace{2mm} \includegraphics[width=0.22\textwidth]{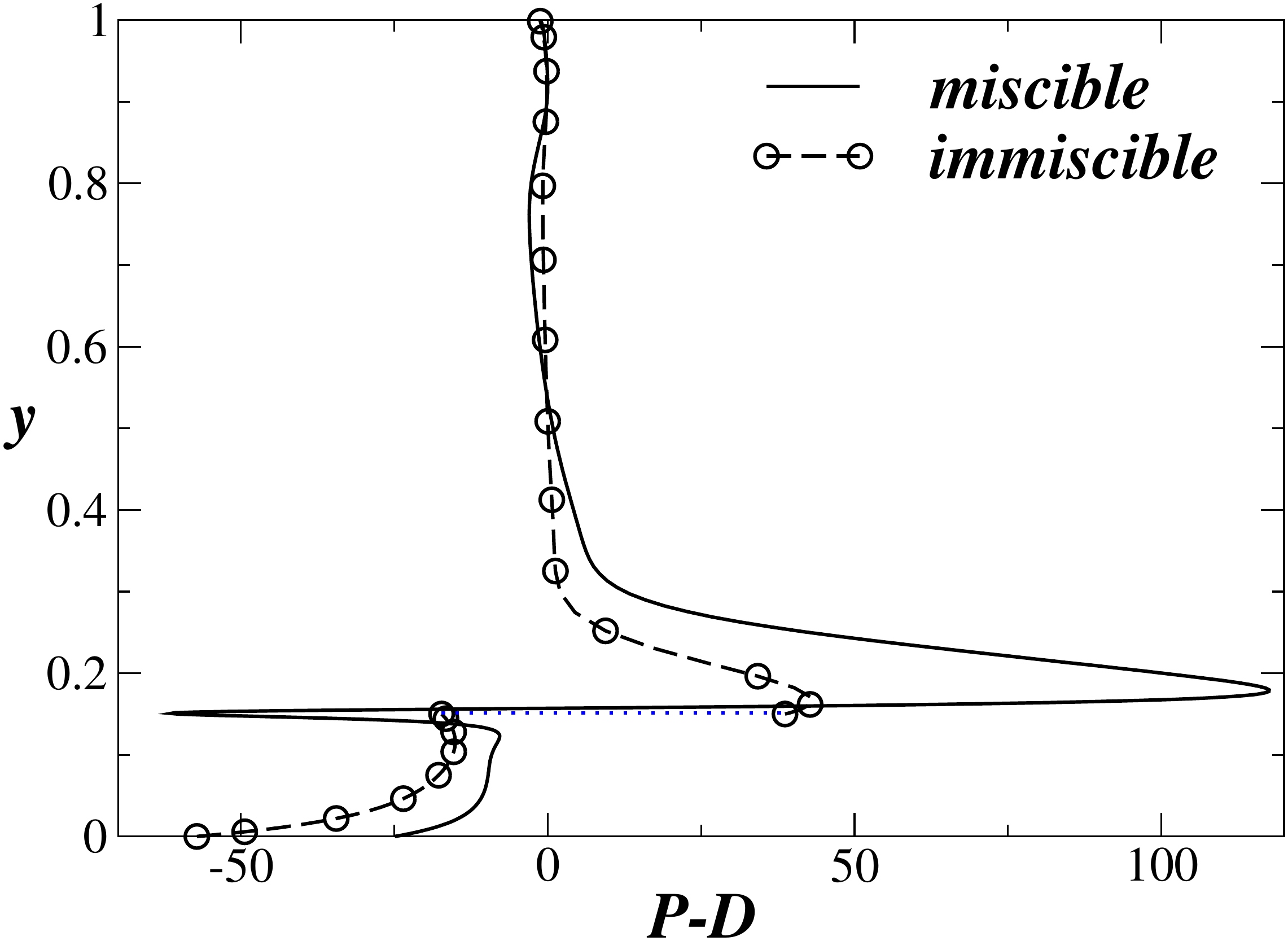}
\caption{Variation of (a)  kinetic energy, $E$ (b) dissipation rate. The negative ($-D$) is shown to help viewing. (c) production rate, $P$ and (d) $P-D$ of the most dangerous disturbance mode ($\alpha=4$) in the wall-normal direction for $\Re=500$. The rest of the parameters are the same as those used to generate Fig. \ref{effectRe}.  The values of $\cal P$, $\cal D$ and $I$ normalised with $\cal E$ for the immiscible case are 0.8458, -1.0745 and 1.1326, respectively; $\omega_{i,max}=0.2159$. For the miscible case: the values of ${\cal P}$ and ${\cal D}$ normalised with ${\cal E}$ are 4.5379 and -1.9227, respectively; $\omega_{i,max}=0.3269$.} 
\label{energy2}
\end{figure}

\subsection{Three-dimensional numerical simulations}

\subsubsection{Numerical method}
\label{sec:num}
For miscible systems, Eqs. (\ref{NS1}) - (\ref{NS3}) are solved by a finite-volume approach \citep{hang2007} using a staggered grid discretization; i.e., the scalar variables (the pressure and concentration of the scalar) and the velocity components are defined at the center and at the cell faces, respectively. The discretized convection-diffusion equation of $s_0$ is given by:
$$
{{{3 \over 2} {s_0}^{n+1}-2 {s_0}^{n} + {1 \over 2}{s_0}^{n-1}} \over \Delta t} = $$
\begin{equation}
{1 \over \Re \Sc} \nabla^2 {s_0}^{n+1}  - 2 \nabla \cdot (\u^n {s_0}^n) + \nabla 
\cdot (\u^{n-1} {s_0}^{n-1}), 
\label{dis1}
\end{equation}
where $\Delta t = t^{n+1}-t^{n}$ and the superscript $n$ represents the time step. The advective terms, i.e. the non-linear terms in Eq. (\ref{NS3}) are discretize using a weighted essentially non-oscillatory (WENO) scheme, and  a central difference scheme is used to discretize the diffusive terms on the right-hand-side of Eqs. (\ref{NS2})-(\ref{NS3}). Second-order accuracy in the temporal discretization is obtained by employing the Adams-Bashforth and the Crank-Nicolson methods for the advective and second-order dissipation terms in Eq. (\ref{NS2}), respectively. This discretized form of Eq. (\ref{NS2}) is given by
$$
{\u^*-\u^n \over \Delta t} = {1 \over p^{n+1/2}} \Big \{-\left[{3 \over 2} 
{\cal H}(\u^n)- {1 \over 2} {\cal H}(\u^{n-1}) \right] + $$
\begin{equation}
{1 \over 2 \Re} \left [{\cal L}(\u^*,\mu^{n+1}) + 
{\cal L} (\u^n,\mu^n) \right]\Big\},
\label{dis3} 
\end{equation}
where $\u^*$ is the intermediate velocity, and ${\cal H}$ and ${\cal L}$ denote the discrete convection and diffusion operators, 
respectively. The intermediate velocity $\u^*$ is then corrected to ${(n+1)}^{th}$ time level.
\begin{equation}
{\u^{n+1}-\u^* \over \Delta t} = {\nabla p^{n+1/2}}.
\label{dis4}
\end{equation} 
The pressure distribution is obtained from the continuity equation at time step ${n+1}$ using
\begin{equation}
\nabla \cdot \left ({\nabla p ^{n+1/2}} \right) = {\nabla \cdot \u^* \over \Delta t}.
\label{dis5}
\end{equation}
For the corresponding immiscible system, a diffuse interface method based on \cite{hang2007} is used. In this case, instead of the diffusion equation (i.e. Eq. \ref{NS3}), the Cahn-Hilliard equation, given by
\begin{equation}
{\partial s_0 \over \partial t} + \u \cdot \nabla s_0 = {1 \over \Re {\Sc}_i} \nabla \cdot (M \nabla \phi),
\label{CH_eq}
\end{equation}
 where ${\Sc}_i \equiv \mu_1 /(\rho M_c \phi_c)$, is solved. Here $M_c$ and $\phi_c$ are the characteristic values of mobility and chemical potential, $\phi$ ($\equiv \epsilon^{-1} \sigma \alpha \Psi^\prime(s_0) - \epsilon \sigma \alpha \nabla^2 s_0$), respectively, wherein $\epsilon$ is the measure of interface thickness, $\Psi(s_0) = {1 \over 4} {s_0}^2 (1-{s_0})^2$ is the bulk energy density, and $\alpha$ is a constant.

\begin{figure}
\centering
\includegraphics[width=0.45\textwidth]{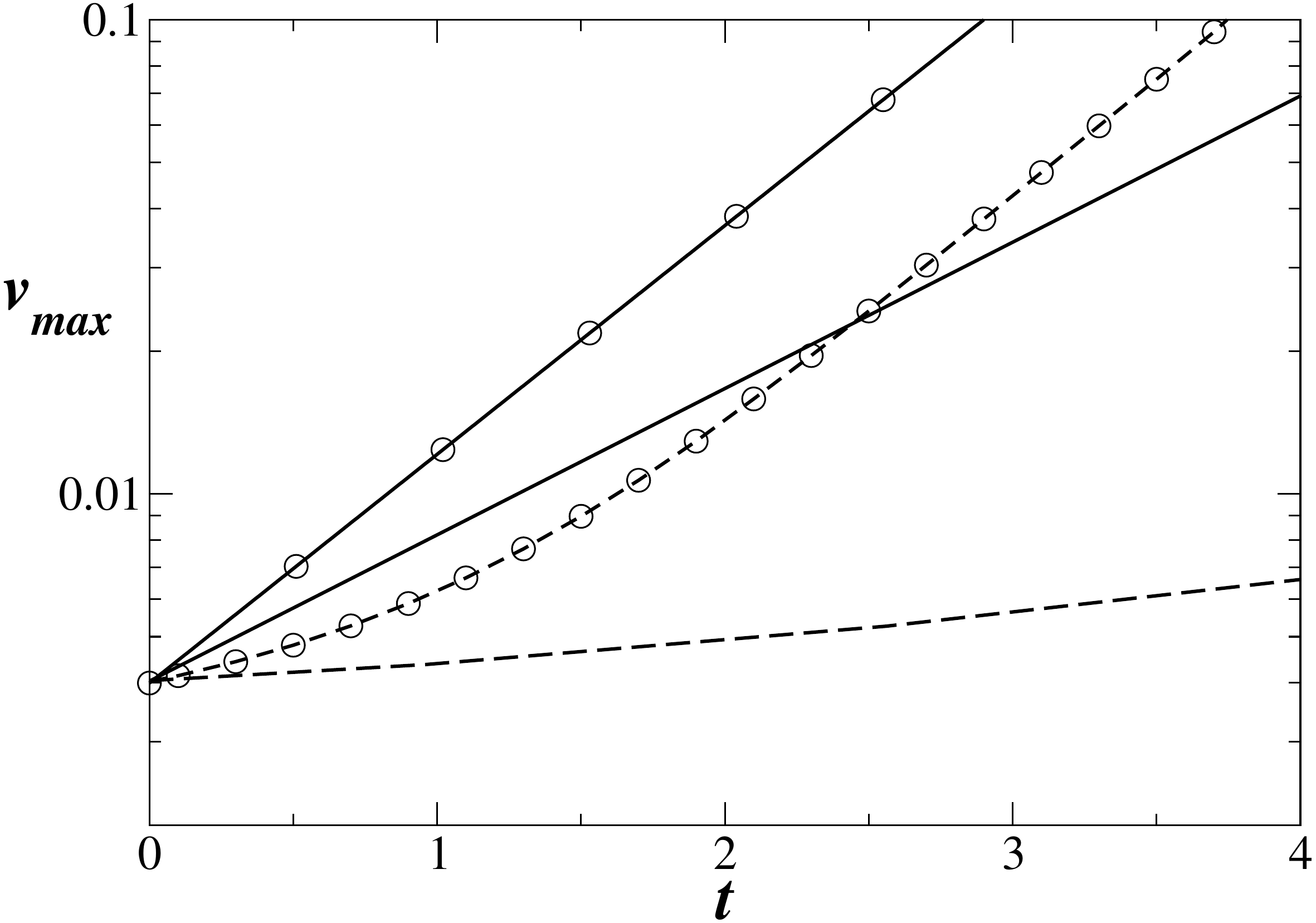}
\caption{Comparison of the temporal evolution of the vertical perturbation obtained from linear stability analysis (solid lines) and direct numerical simulations (dashed lines). Miscible flow: symbols, Immiscible flow: without symbols. The rest of the parameter values are $\Re=500$, $\Sc=100$, $R_s=2$, $h=0.3$ and $q=0.01$.}
 \label{comp}
\end{figure}

{The above discretized equations are solved by employing the no-slip and the no-penetration boundary conditions for the velocity components and no-flux condition for concentration $(s_0)$ at the walls, and periodic boundary conditions in the axial and lateral directions. The pressure gradient is kept the same as the that of the stability analysis conducted in the previous section. A domain with 321, 81, and 161 cells in the axial $x$, wall-normal $y$ and spanwise $z$, directions, respectively are used in the simulations.} Grid refinement tests have been conducted to ensure that we obtain grid-converged results. The numerical procedure described above for miscible and immiscible systems are similar to the ones used by \cite{hang2007}, respectively. The reader is referred to this paper for detailed descriptions and validation of the solvers. 

\begin{figure}
\centering
(a) \hspace{4cm} (b) \\
\includegraphics[width=0.5\textwidth]{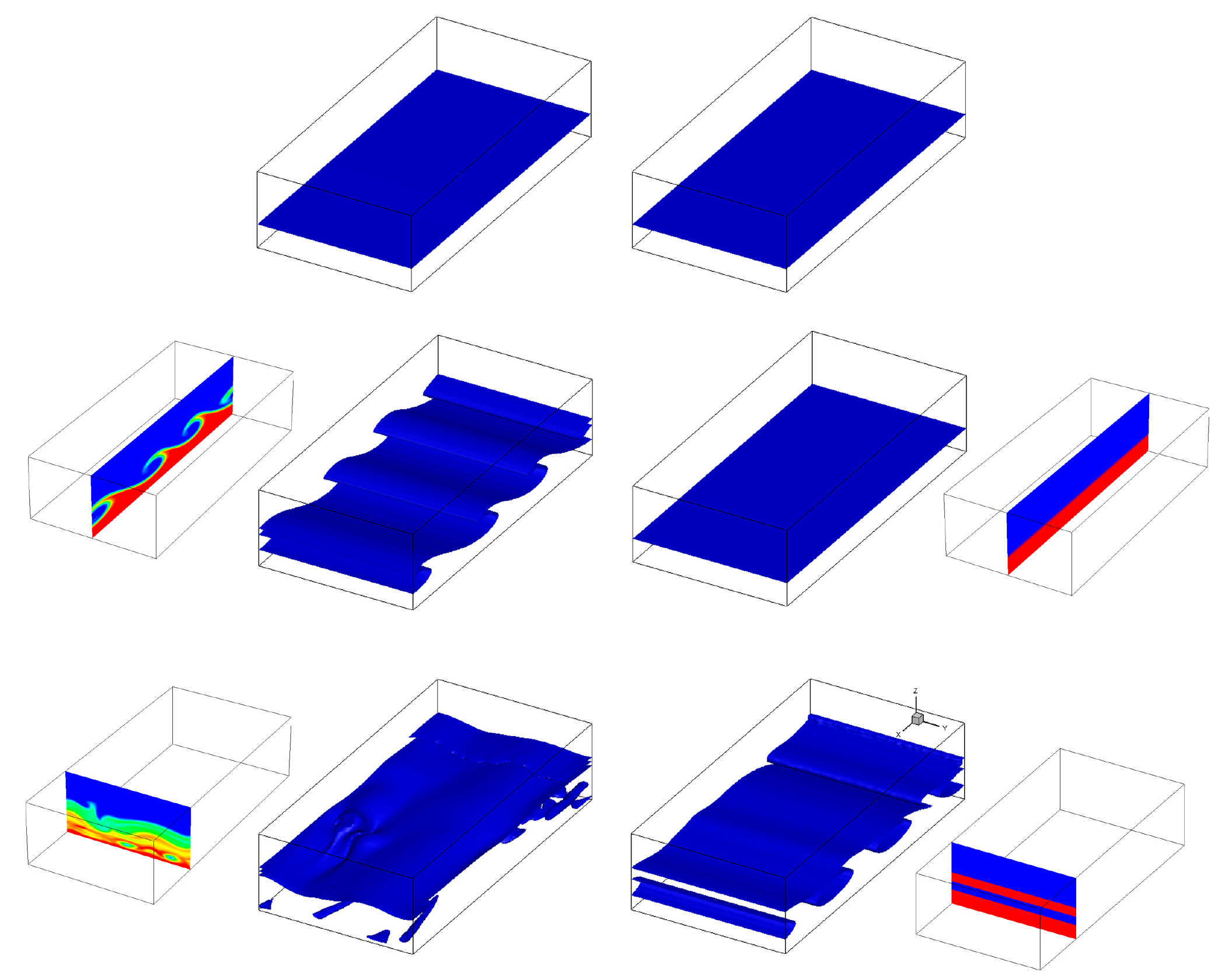}
\caption{Temporal evolution of (a) $s=0.5$ contour for miscible flow with $q=0.05$, $\Sc=100$, (b) interface for immiscible case. From top to bottom: $t=4$, $6$, and $10$. The rest of the parameter values are $h=0.3$, $\Re=500$ and $R_s=2$. The flow is along the positive $x$ direction. The side panels at $t=6$ are the cross-sectional views in the $x$-$z$ plane at $y=1$, and those at $t=10$ are the cross-sectional views in the $y$-$z$ plane at $x=3$} \label{dns3}
\end{figure}

\begin{figure}
\centering
(a) \hspace{4cm} (b) \\
\includegraphics[width=0.5\textwidth]{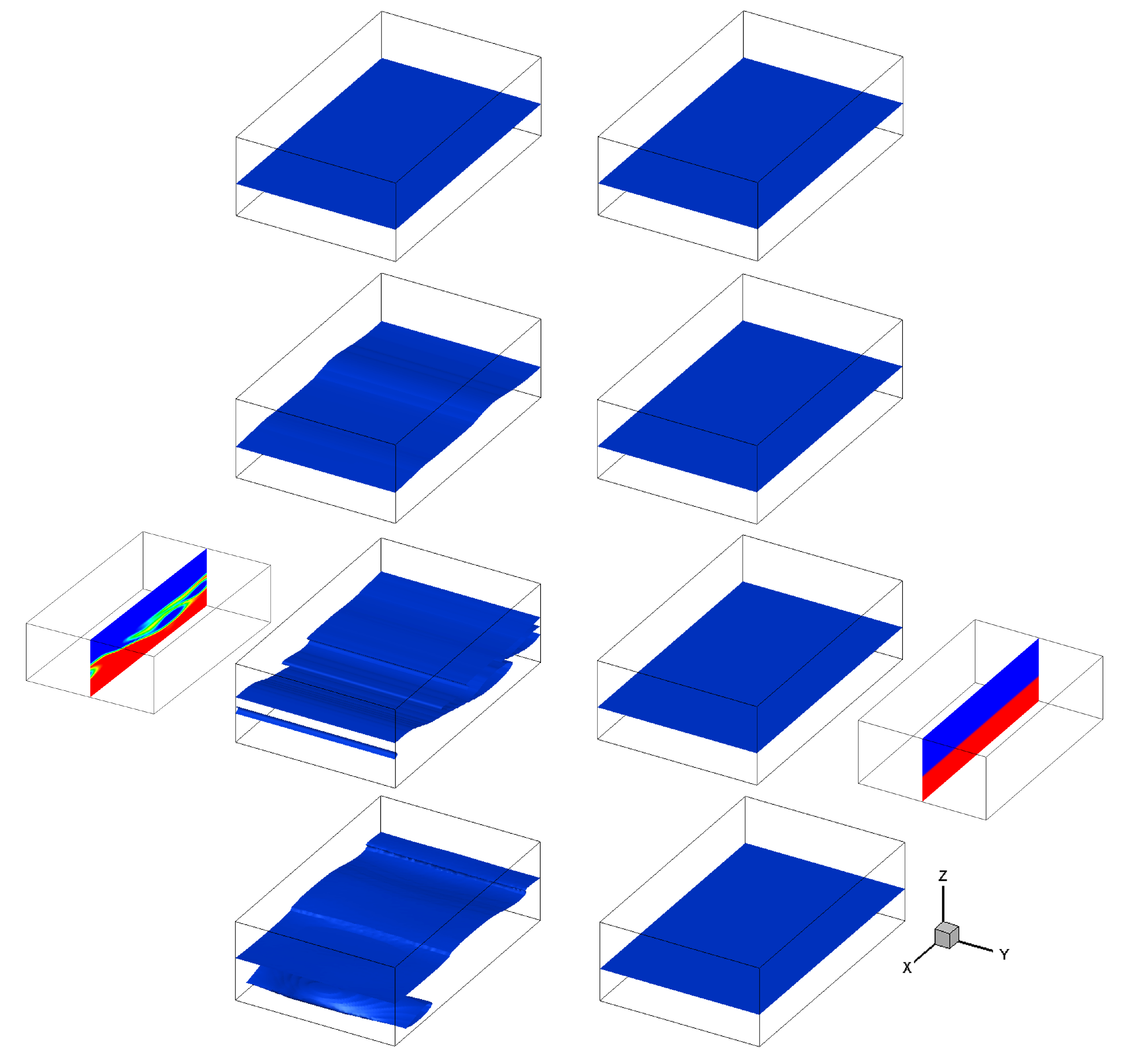}
\caption{Temporal evolution of (a) $s=0.5$ contour for miscible flow with $q=0.01$, $\Sc=100$, (b) interface for immiscible case. From top to bottom: $t=3$, $5$, $7$ and $9$. The rest of the parameter values are $h=0.4$, $\Re=500$ and $R_s=2$. The flow is along the positive $x$ direction. The insets at $t=7$ represent the cross-sectional viewes in the $x$-$z$ plane.} \label{dns2}
\end{figure}

In Fig. \ref{comp}, we compare the maximum value of the wall-normal velocity component, $v_{max}$ obtained from our direct numerical simulations with that obtained from linear stability analysis for $\Re=500$, $\Sc=100$, $R_s=2$, $h=0.3$ and $q=0.01$. The miscible and immiscible simulations are conducted inside a channel whose length is twice the wavelength of the most-dangerous mode obtained from linear stability analysis ($\alpha=4$ for this set of parameter values). If we were to prescribe the dominant perturbation mode in the initial conditions, we would expect this perturbation to grow in consonance with linear theory at early times. We choose to study the harder case, i.e., one in which we prescribe no initial perturbation. Thus all modes of perturbation are equally initialised, so the dominant one will need time to become visible and to grow in accordance with linear theory. In the miscible case, we see that in some time (after $t=1$), the growth rate of the disturbance in the numerical simulations is close to that of the dominant mode predicted by linear theory (seen by the fact that the two lines are parallel). In the immiscible case, the dominant mode is not distinguishable by its linear growth rate at any time, indicating that nonlinear effects dominate the entire process. In the nonlinear regime too, it can be observed that the miscible flow with finite Schmidt number is more unstable than the immiscible flow. The subsequent three-dimensional plots (Figs. \ref{dns3} and \ref{dns2}) convey the same information but in a pictorial form of an iso-concentration contour.

A few cases were computed, of which we present results for a Reynolds number of $\Re=500$ and $R_s=2$ as being representative. The spatio-temporal evolution of the interface separating the fluids in the immiscible flow ($q=0$, $\Sc = \infty$) and the $s=0.5$ contour for the miscible flow ($\Sc=100$) for $q=0.01$, $h=0.4$ and $q=0.05$, $h=0.3$ are presented in Figs. \ref{dns3} and \ref{dns2}, respectively. A computational domain of size $4 \times 2 \times 1$ is used for these simulations, wherein velocity components are set to zero initially. Given constraints of numerical accuracy, we did not perform simulations with thinner mixed layers. It can be seen in Fig. \ref{dns3}(a) that the miscible flow becomes unstable, and rolled-up structures are obtained, which can be observed at $t=6$ in the contour of $s=0.5$. At later times the nonlinear instability develops an irregular lateral structure too (see the side panel at $t=10$ in Fig. \ref{dns3}(a)). However, it can be seen in Fig. \ref{dns3}(b) that the interfacial flow becomes unstable at a time much later than the corresponding miscible flow. Secondly the immisicble flow remains two-dimensional in the regime where the miscible flow has become three-dimensional. For $h=0.4$ (Fig. \ref{dns2}) the contrast is even more pronounced; in this case the interfacial flow is stable till $t=9$, whereas the miscible flow becomes unstable at $t \approx 5$. At early times this behaviour is consistent with that obtained in the linear stability analysis, and at later times it is seen that the tendency for the immiscible flow to remain less unstable than the miscible persists into the nonlinear regime as well. 

\section{Discussion and summary}
\label{sec:conclusion}
In the present study, we have investigated the difference between interfacial flow instabilities {(without surface tension)} with a viscosity jump across the interface, and instabilities associated with two-layer miscible channel flow of two fluids with different viscosities. We show that in this flow too, an overlap mode of instability, seen before in other miscible flow configurations \citep{rg2014} is dominant, where the mixed layer overlaps significantly with the critical layer of the dominant disturbance. This means that the lower order terms in the critical layer balance get disturbed by the viscosity variation in the mixed layer and contribute to significant changes in the stability behaviour. The overlap mode of instability is produced by an inertial effect, and is driven by different physics from instabilities seen before in this flow under zero Reynolds number conditions. In fact the overlap mode of instability will necessarily vanish at zero Reynolds number. At Reynolds numbers of order $1$ or lower, the critical layer is as wide as the flow, and is thus unable to produce singular effects. Interestingly the overlap mode of instability becomes operational at relatively low Schmidt numbers and low viscosity ratios, unlike the mechanisms in operation at very low Reynolds number, which are triggered by high viscosity ratio and poor diffusivity.

We have studied how decreasing the diffusivity of the fluids takes the results closer to the immiscible case. Above a Schmidt number of $\sim 10^5$, the behaviour of the miscible layer is very close to that of the immiscible. At moderate Schmidt number, of up to $100$, we find that the channel flow of two miscible fluids can be more unstable than the case where the two fluids are immiscible. The increase in the disturbance growth rate due to finite diffusivity is not too large, but makes a point of principle which needs further investigation. {Whether this effect at high Reynolds number has any connection with the destabilisation due to diffusivity seen at very low Reynolds numbers by \cite{ern03a} is not established yet,} but as discussed above we have strong reasons to believe that the two are completely different. 

As expected, increasing the Reynolds number destabilises the flow. We find that increasing the viscosity contrast between the two fluids does not have any significant effect on instability characteristics in the immiscible case, but significantly increases the growth rate for the miscible two-layer flow. Reducing the thickness of the mixed layer increases the growth rate, as expected. Thus for thin mixed layers at intermediate diffusivity, the increase in instability as compared to the immiscible case is larger Varying the location of the mixed layer or the interface has an effect on the stability. In some range of this parameter, distinct regions of overlap instability are obtained, whereas in others, the instability regions due to different mechanisms merge with each other.

The miscible two-layer channel flow had not been studied earlier in the nonlinear regime to our knowledge, and we therefore conduct direct numerical simulations for both the miscible and immiscible cases. At a Schmidt number of $100$, linear stability analysis predicts a faster growth rate for miscible than for the immiscible. This is borne out by the simulations. Also we see that nonlinear effects on the immiscible flow are visible at earlier times than in the miscible, and the rate at which we see the interface roll-up can be made much slower by immiscibility, or even suppressed.

\section*{{Appendix: Validity of the parallel flow assumption}}

Consider a situation when a splitter plate is located at $x < x_0$, at a constant $y$ and parallel streams of two miscible fluids flow on both sides of this plate. The streams come into contact with each other at $x = x_0$. The two fluids begin to mix with each other for $ x > x_0$, thus producing a stratified layer. The thickness `$q$' of this layer grows as the fluids move in the downstream direction and therefore $q$ is a function of $x$. We note that the flow diffuses as it moves downstream but does not diffuse in time at one $x$ location, i.e., the base flow is steady in time.

We know that at any location, the concentration $s_0$ satisfies the following equation,
\begin{equation}
\frac{\partial s_0}{\partial t} + U \frac{\partial s_0}{\partial x} + V \frac{\partial s_0}{\partial y} = \frac{1}{\Re \Sc}
\left [ \frac{\partial^2 s_0}{\partial x^2} + \frac{\partial^2 s_0}{\partial y^2}  \right ], 
\label {a12}
\end{equation}
For slow diffusion (i.e for high $Pe \equiv \Re \Sc$), we can make the assumption on locally parallel flow (variation of $s_0$ in the $y$ direction is much larger than that in the $x$ direction); thus $V \ll U$ and $\frac{\partial^2}{\partial x^2} \ll \frac{\partial^2}{\partial y^2}$. This is equivalent to saying that the variations of the gradients of the flow variables and the thickness $q$ of the mixed region have much larger length scale than the disturbance wavelength. In such a scenario, the concentration is a function of $y$ and $t$ only and not of $x$. Thus the above equation reduces to 
\begin{equation}
\frac{\partial s_0}{\partial t} = \frac{1}{Pe}\frac{\partial^2 s_0}{\partial y^2}.
\label{eq1}
\end{equation}
Using the same approximation, we know that $U \sim O(1)$,  $y \sim \sqrt{\nu}$, where $\nu$ is the kinematic viscosity. As the viscosity is directly proportional to the concentration in the mixed layer. Therefore, $q s_0 \sim O(y^2)$. This implies that ${\partial s_0 / \partial x} \simeq \frac{1}{q}\,O({1/Pe})$. Thus for $Pe>>1$, ${\partial s_0 / \partial x}$ is very small, i.e., the downstream variation of $s_0$ is very small which in turn implies that the change in the thickness of the mixed layer $(q)$ along the $x$-direction is very small.

Alternatively, if we assume a similarity solution $s_0 (y/q(x)) \simeq s(\xi)$ (where $\xi = (y/q(x))$), from equation (1), we get
\begin{equation}
U \frac{ds_0}{d\xi}\left (- \frac{\xi}{q}\,\frac{dq}{dx} \right ) \simeq \frac{1}{Pe} \left (\frac{d^2 s_0}{d\xi^2}\,\frac{1}{q^2} \right). 
\label {a13}
\end{equation}
As a consequence,
\begin{equation}
\frac{1}{q}\,\frac{dq}{dx} \sim \frac{1}{q^2 Pe} \Rightarrow \frac{dq}{dx} \sim \frac{1}{q}\, O(Pe)^{-1}.
\label {a14}
\end{equation}
Thus, the downstream growth of mixed layer is inversely proportional to the P\'{e}clet number as $U$ and $\xi$ are of $O(1)$, and $O(\frac{ds_0}{d\xi})\simeq O(\frac{d^2s_0}{d\xi^2})$. For most of the simulations considered in the present study, $Pe \ge 1000$, $\Re \ge 100$ and $q \ge 0.05$; these parameter values are well above the limit for which parallel flow assumption is valid. This confirms that for the Reynolds and the Schmidt numbers considered in the present study,  the assumption of uniform thickness of viscosity stratified layer is justified.

\begin{figure}
\centering
(a)\\
\includegraphics[width=0.45\textwidth]{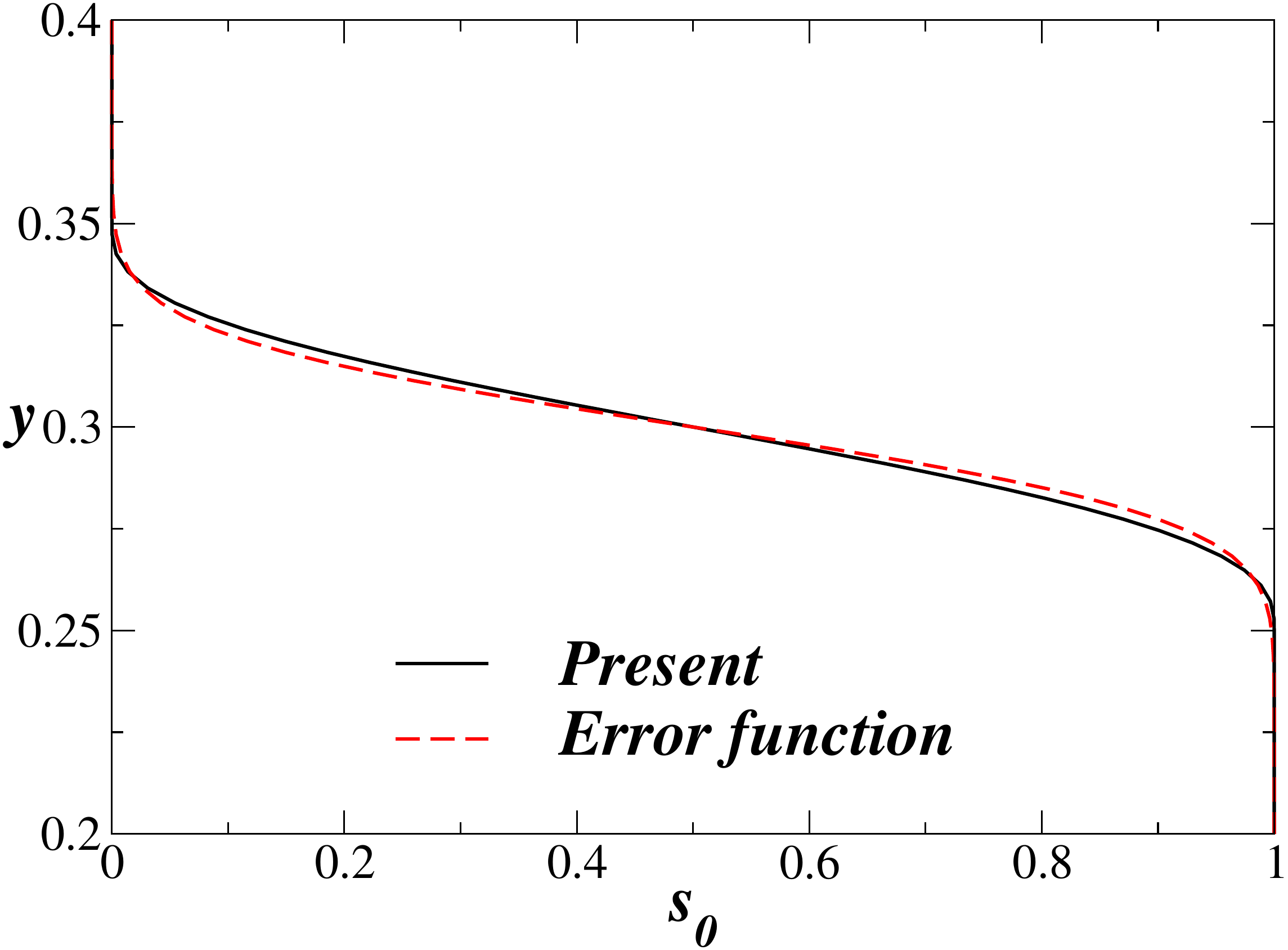} \\
 (b) \\
 \includegraphics[width=0.45\textwidth]{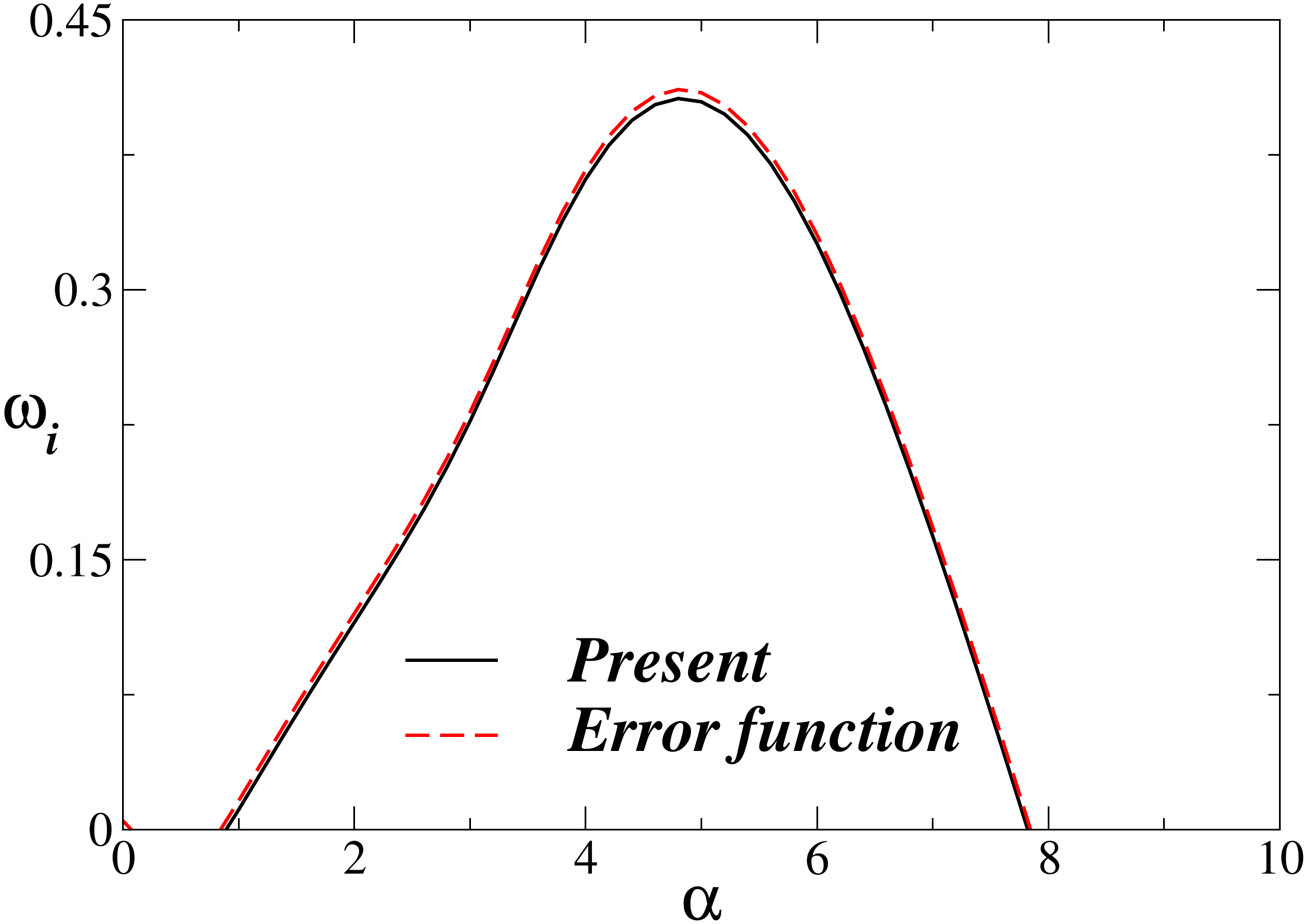}\\
\caption{(a) $S_0$ profile, (b) the dispersion curve ($\omega_i$ versus $\alpha$) obtained using the present base state and an error function profile. The other parameters are chosen as $\Re=500$, $\Sc=10$, $R_s=1$, $h=0.3$ and $q=0.1$.} 
\label{fig12}
\end{figure}

{Now, the solution of Eq. (\ref{eq1}) is an error function, and the fifth order polynomial is a good representation of this, as seen in Fig. \ref{fig12}(a) in this response. The growth rates obtained using an error function type profile and the fifth order polynomial used in the present study are compared in Fig. \ref{fig12}(b). It can be seen that they too agree very well.}

{We also note that such basic flows are commonly used in stability studies. Several authors have used the same logic to give a basic concentration profile in the form of a hyperbolic tangent \citep{ern03a} or an error function (e.g. \cite{talon11a,selvam09a}), given by
\begin{equation}
s_0= 0.5 - 0.5 \erf \left [{{y-h-0.5q} \over q} \right ].
\end{equation}
Some have also used a fifth-order polynomial (see e.g. \cite{malik05a}), which is smooth enough to approximate either profile.}

In summary, the disturbance wavelength is much shorter than the downstream length scale over which the mixed-layer thickness registers any growth, so it is justified to use locally a constant-thickness approximation.

\section*{Acknowledgement}
The authors would like to sincerely thank the anonymous reviewers for their valuable comments and suggestions.


\end{document}